\documentclass[aps,pra,twocolumn,preprintnumbers,nofootinbib,amsmath,amssymb,superscriptaddress]{revtex4-2}

\usepackage{amsmath}
\usepackage{float}
\usepackage{amsfonts}
\usepackage{amssymb}
\usepackage{subfiles}
\usepackage{natbib}
\usepackage[colorlinks,linkcolor=red,citecolor=blue,breaklinks=true]{hyperref}
\usepackage{graphicx}
\usepackage{dcolumn}
\usepackage{bm}
\usepackage{color}
\usepackage{booktabs}
\usepackage{amsthm}
\usepackage{dsfont}
\usepackage{tcolorbox}
\usepackage{cases}
\usepackage{physics}
\usepackage{subfiles}

\usepackage{braket}
\usepackage{bbold}
\usepackage{xcolor}
\usepackage[normalem]{ulem}
\usepackage{txfonts,times}
\usepackage{orcidlink}
\usepackage{lipsum}
\usepackage{tabularx}
\usepackage{booktabs}

\newcommand{\eqnumreset}{\setcounter{equation}{0}}

\begin{document}

\title{Quantum switch instabilities with an open control}

\author{Otavio A. D. Molitor\orcidlink{0000-0002-8582-5751}}
\email{dmolitor.oa@protonmail.com}
\address{International Centre for Theory of Quantum Technologies, University of Gdańsk, Jana Bażyńskiego 1A, 80-309 Gdańsk, Poland}

\author{André H. A. Malavazi\orcidlink{0000-0002-0280-0621}}
\email{andrehamalavazi@gmail.com}
\address{International Centre for Theory of Quantum Technologies, University of Gdańsk, Jana Bażyńskiego 1A, 80-309 Gdańsk, Poland}

\author{Roberto Dobal Baldij\~{a}o\orcidlink{0000-0002-4248-2015}}
\email{roberto.dobal-baldijao@ug.edu.pl}
\address{International Centre for Theory of Quantum Technologies, University of Gdańsk, Jana Bażyńskiego 1A, 80-309 Gdańsk, Poland}

\author{Alexandre C. Orthey Jr.\orcidlink{0000-0001-8111-3944}}
\email{aorthey@cft.edu.pl}
\affiliation{Center for Theoretical Physics, Polish Academy of Sciences, Al. Lotnik\'ow 32/46, 02-668 Warsaw, Poland.}

\author{Ismael L. Paiva\orcidlink{0000-0002-0416-3582}}
\email{ismaellpaiva@gmail.com}
\affiliation{H. H. Wills Physics Laboratory, University of Bristol, Tyndall Avenue, Bristol BS8 1TL, United Kingdom}

\author{Pedro R. Dieguez\orcidlink{0000-0002-8286-2645}}
\email{dieguez.pr@gmail.com}
\affiliation{International Centre for Theory of Quantum Technologies, University of Gdańsk, Jana Bażyńskiego 1A, 80-309 Gdańsk, Poland}

\begin{abstract}
The superposition of causal order shows promise in various quantum technologies. However, the fragility of quantum systems arising from environmental interactions, leading to dissipative behavior and irreversibility, demands a deeper understanding of the possible instabilities in the coherent control of causal orders. In this work, we employ a collisional model to investigate the impact of an open control system on the generation of interference between two causal orders.  We present the environmental instabilities for the switch of two arbitrary quantum operations and examine the influence of environmental temperature on each potential outcome of control post-selection. Additionally, we explore how environmental instabilities affect protocol performance, including switching between mutually unbiased measurement observables and refrigeration powered by causal order superposition, providing insights into broader implications.
\end{abstract}

\maketitle

\section{Introduction}

\begin{figure*}
    \centering
    \includegraphics[width=\textwidth]{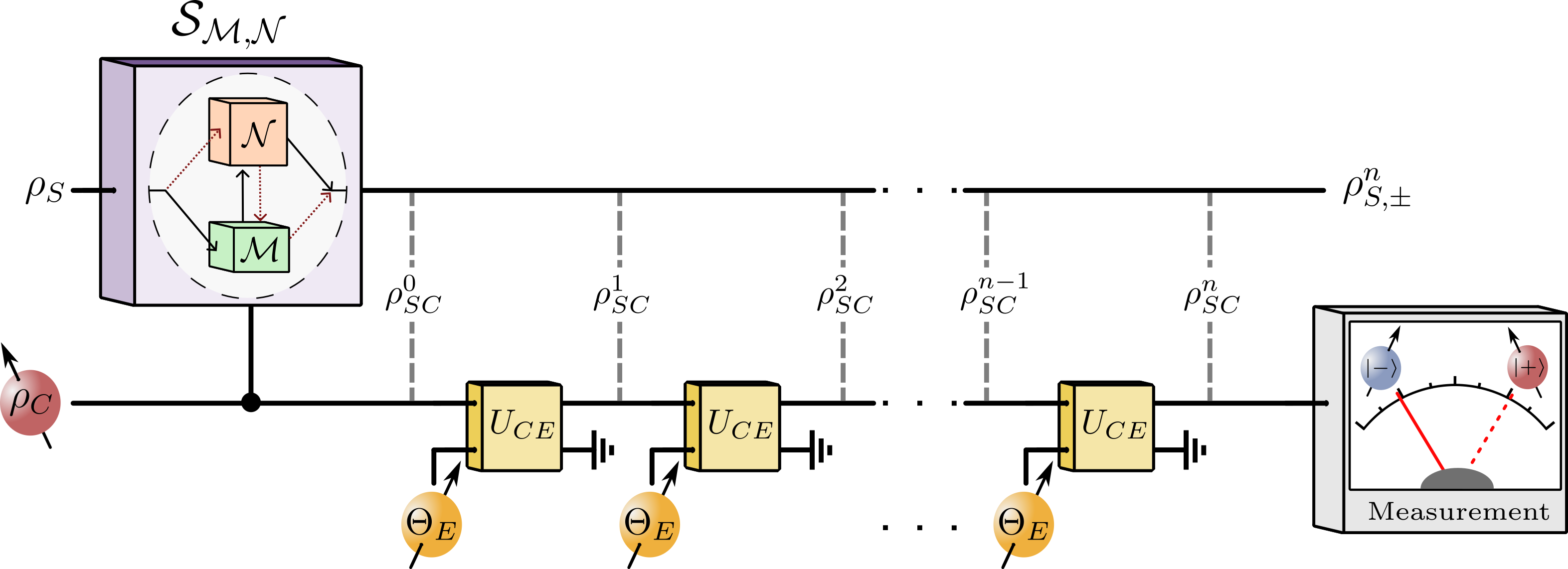}
    \caption{\textbf{General scheme of the model analyzed in this work.} One starts with a system in the state $\rho_S$ and a control in the state $\rho_C$. Initially, they are uncorrelated and the composite state $\rho_S\otimes\rho_C$ undergoes a quantum switch (QS) map $\mathcal{S}_{\mathcal{M},\mathcal{N}}$. Before being measured, the control is then considered to interact with an external thermal environment with the inverse of temperature $\beta_E$, which is modeled according to the collisional model framework. The environment then consists of a stream of qubits in the thermal state $\Theta_E = \text{exp}(-\beta_E H_E)/Z_E$ ($H_E$ and $Z_E=\tr\{\text{exp}(-\beta_E H_E)\}$ are the Hamiltonian and partition function, respectively) which one-by-one come and \textit{locally} interact with the control degree-of-freedom through the unitary $U_{CE}$. After each collision, the new composite system-control state is given by $\rho_{SC}^k$, where $k$ corresponds to the number of collisions that have already happened. Finally, after $n$ collisions, one measures the $\sigma_x$ operator of the control, which is equivalent to making projections onto the basis $\{\ket{+}_C, \ket{-}_C\}$. The result of the measurement directs the local state of the system to either $\rho_{S,+}^n$ or $\rho_{S,-}^n$.}
    \label{fig:generalscheme}
\end{figure*}

Quantum coherence is one of the notable features of the quantum description of nature that distinguishes it from classical theories~\cite{baumgratz2014quantifying, streltsov2017colloquium, wu2020quantum, designolle2021set, dieguez2022experimental, giordani2023experimental}. This intrinsically quantum phenomenon can be employed to lead to the indefiniteness of the causal structures underlying the application of quantum operations~\cite{hardy2005probability, oreshkov2012quantum, oreshkov2019time}, which may be a resource for new quantum advantages~\cite{milz2022resource}. Paradigmatic examples associated with this are given by processes that utilize an auxiliary quantum control to devise a superposition of causal order (SCO) of operations in a system of interest, such as the \textit{quantum switch} (QS)~\cite{chiribella2013quantum}. SCO has been interpreted as a superposition of time evolution~\cite{felce2022indefinite} and applied in a plethora of fields~\cite{rozema2024experimental}, ranging from computation~\cite{chiribella2013quantum, araujo2014computational, procopio2015experimental} and communication~\cite{ebler2018enhanced, wei2019experimental, guo2020experimental} to metrology~\cite{zhao2020quantum} and thermodynamics~\cite{Felce2020, rubino2021quantum, nie2022experimental, Simonov2022, dieguez2023thermal, Simonov2023}. However, there are still open questions and a debate about whether (and how) the SCO resulting from the QS (or other processes with a quantum control) can be considered genuine indefinite causal order and benefit from the advantages associated with it~\cite{goswami2018indefinite, purves2021quantum, capela2023reassessing}.

The quantum correlations formed when applying a QS are essential to observe SCO effects after the control post-selection procedure~\cite{chiribella2013quantum, goswami2018indefinite}. Therefore, it is imperative to understand and characterize their resilience under more realistic scenarios as it is widely recognized that quantum states are fragile due to their unavoidable interaction with environmental degrees of freedom~\cite{zurek2003decoherence, schlosshauer2005decoherence}. In general, these interactions lead to non-unitary processes accompanied by dissipation and irreversibility, which directly affects the existence of quantum resources and idealized closed dynamical frameworks fail to capture. The microscopic derivation of open quantum system dynamics is realized by explicitly including and modeling the external environment. However, this process can become intricate when dealing with more general dynamics that extend beyond the standard regime defined by the weak coupling, Born--Markov, and secular approximations~\cite{Breuer, rivas2012open}. A compelling alternative is given by the \textit{collisional model} framework~\cite{Scarani2002, Ziman2002, Karevski2009, Giovannetti2012, Landi2014, Barra2015, Strasberg2017, DeChiara2018, Rodrigues2019, Molitor2020}, which consists of modeling the environment as a set of identically prepared auxiliary systems interacting with the system of interest through some unitary evolution. Despite its straightforward conceptual and procedural nature, this allows one to approach broad physical scenarios. In this sense, one can choose the initial state of the auxiliary systems (e.g., Gibbs states for thermal reservoirs) and the interaction terms, consider non-Markovianity~\cite{ciccarello2022quantum}, and derive local master equations under an appropriate scaling of the interaction strength~\cite{DeChiara2018}.

In this work, we recognize the post-selection of the control as crucial for SCO effects in the QS. Yet it also poses a potential exposure of the control to environmental interactions. To account for such a process,
we thus employ the collisional model to examine the robustness of correlations between the target and control states before the post-selection procedure in the QS of two \textit{arbitrary} maps. The proposed general procedure is elucidated by considering the reservoir auxiliary systems as a set of qubits in the Gibbs state, coupling with the system of interest through an excitation-conserving interaction. Such an approach provides analytical results for the QS with an open control. Within this analysis, we detect thermal instabilities caused by environmental interactions. Remarkably, the instabilities in the quantum switch of arbitrary quantum operations consistently diminish the contribution of SCO terms, independently of the environment temperature. However, the temperature influences the post-selection probabilities and conditional states. In the low-temperature regime, the environment asymmetrically shields one post-selection outcome, while in the high-temperature regime, both outcomes are similarly affected, suppressing SCO. Overall, these findings suggest that environmental interactions qualitatively alter the QS behavior.

To illustrate our findings, we consider two paradigmatic applications. First, we utilize our open-control QS model to analyze the SCO of channels that describe non-selective measurements of incompatible observables, transitioning from weak to strong projective measurement regimes~\cite{oreshkov2005weak, dieguez2018information, dieguez2018weak}. Additionally, we discuss how the dynamics of an open control can impact the coefficient of performance of an SCO-powered refrigerator~\cite{felce2020quantum}. In both cases, the instabilities and asymmetry caused by environmental interactions strongly influence the effectiveness of the QS for the intended application. Throughout the work, we use units such that $\hbar = k_B = 1$.

The remainder of this article is structured as follows. In Sec.~\ref{sec:qs} we introduce the QS setup. In Sec.~\ref{sec:results} we present our model for an open QS and our main results. In Sec.~\ref{sec:discussion}, we discuss our results, connecting them with thermodynamical concepts and presenting an outlook for future research avenues. Finally, we introduce two examples of concrete applications of our results in the Appendices.

\section{Quantum switch}
\label{sec:qs}

Consider a quantum system $S$ initially in a state $\rho_S$ with local Hamiltonian $H_S$. The introduction of an auxiliary control degree-of-freedom $C$ enables the implementation of the QS, where the state $\rho_C$ determines the order of application of two (or possibly more) quantum maps. That is, depending on the state of the control, the maps are applied in different orderings. Given the completely positive trace-preserving (CPTP) maps $\mathcal{M}$ and $\mathcal{N}$ with Kraus operators $\{M_i\}$ and $\{N_i\}$, respectively, such that\footnote{The maps $\mathcal{M}$ and $\mathcal{N}$ act on the same system $S$, which is undergoing the switch. Therefore, they both sum to the identity in the same space, i.e., $\mathds{1}_S$. Note, however, that $i$ and $j$ can run over any finite number of values, which need not coincide.} $\sum_i M_i^\dagger M_i = \sum_i N_j^\dagger N_j = \mathds{1}_S$, the controlled-Kraus operators are written as
\begin{equation}
\label{eq:Wij}
    W_{ij} \coloneqq M_i N_j \otimes |0\rangle \langle0|_C + N_j M_i \otimes |1\rangle\langle 1|_C.
\end{equation}
Thus the effect of the QS map on the system-control state is
\begin{equation}
   \rho_{SC}^{\mathcal{M} \leftrightarrow \mathcal{N}} \coloneqq \mathcal{S}_{\mathcal{M},\mathcal{N}}(\rho_S \otimes \rho_C) = \sum_{ij} W_{ij}(\rho_S \otimes \rho_C) W_{ij}^{\dagger},
   \label{eq:switch}
\end{equation}
where the composite system-control state is assumed to be initially separable. It follows that, if the control is in the state $\ket{0}_C$ or $\ket{1}_C$, the maps are applied in the definite order characterized by the sequential application of $\mathcal{M}$ and $\mathcal{N}$ or vice-versa, respectively. Hence, if the control is in a \textit{coherent state}, e.g., $\ket{\psi}_C = \sqrt{p} \ket{0}_C + \sqrt{1-p}\ket{1}_C$ ($p \neq 0,1$), a superposition between the causal orders can be achieved. Note that two states of $C$ are sufficient to implement a QS between two processes, which allows one to effectively model the control as a two-level system (i.e., a qubit) regardless of the dimension of $S$. Then, the composite system-control state post-QS can be written as~\cite{Simonov2023}
\begin{equation}
\label{eq:statepostQSSimonov}
    \begin{aligned}
    \rho_{SC}^{\mathcal{M}\leftrightarrow\mathcal{N}}= &\; A_{++}\otimes\rho_{C}+A_{+-}\otimes\rho_{C}\sigma_{z}\\
    &+A_{-+}\otimes\sigma_{z}\rho_{C}+A_{--}\otimes\sigma_{z}\rho_{C}\sigma_{z},
    \end{aligned}
\end{equation}
where $\sigma_z$ is the $z$-Pauli matrix and we have defined the operators
\begin{equation}
    \label{eq: Axy}
    A_{xy} \coloneqq \frac{1}{4}\sum_{i,j}\left[M_{i},N_{j}\right]_{x}\rho_{S}\left[M_{i},N_{j}\right]_{y}^{\dagger}
\end{equation}
with $x,y\in\{+,-\}$, $[X,Y]_{-} \coloneqq XY - YX$ (i.e., the commutator) and $[X,Y]_{+} \coloneqq XY + YX$ (i.e., the anti-commutator). Eq.~\eqref{eq:statepostQSSimonov} is fully general regarding the channels applied in the QS and the initial state of the control. For the purposes of this work, however, we take the paradigmatic special case of the initial state of the control being $\rho_C=|+\rangle \langle+|_C$, which features maximal coherence in the computational basis and, therefore, is among the best suitable initial states to explore SCO. Thus, Eq.~\eqref{eq:statepostQSSimonov} becomes
\begin{align}
\label{eq:SCstate_+input}
    \rho_{SC}^{\mathcal{M}\leftrightarrow\mathcal{N}} = &\; A_{++}\otimes\ketbra{+}{+}+A_{+-}\otimes\ketbra{+}{-}\\
    &+ A_{-+}\otimes\ketbra{-}{+}+A_{--}\otimes\ketbra{-}{-}.\nonumber
\end{align}
The joint state $\rho_{SC}^{\mathcal{M} \leftrightarrow \mathcal{N}}$ carries terms related to SCO. To see that, define the following operators
\begin{equation}
\label{eq:A_def}
    A_{\textrm{def}} \coloneqq A_{++}+A_{--} = \frac{1}{2}\sum_{i,j}\left(M_{i}N_{j}\rho_{S}N_{j}^{\dagger}M_{i}^{\dagger}+N_{j}M_{i}\rho_{S}M_{i}^{\dagger}N_{j}^{\dagger}\right)
\end{equation}
and
\begin{equation}
\label{eq: A_indef}
    A_{\textrm{indef}} \coloneqq A_{++}-A_{--} = \frac{1}{2}\sum_{i,j}\left(M_{i}N_{j}\rho_{S}M_{i}^{\dagger}N_{j}^{\dagger}+N_{j}M_{i}\rho_{S}N_{j}^{\dagger}M_{i}^{\dagger}\right).
\end{equation}
Observe that $A_{\rm def}$ is a convex combination of two terms: One with $\mathcal{M}$ applied to the system, followed by $\mathcal{N}$, and the other term representing the opposite order. Therefore, $A_{\rm def}$ corresponds to a mixture of definite orders. $A_{\textrm{indef}}$, however, corresponds to interference terms between the causal orders, i.e., terms without definite causal order in the quantum description. Since 
\begin{align}
\label{eq:ApmpmIntermsofAdefindef}
    A_{\pm\pm}=\frac{1}{2}A_{\rm def}\pm \frac{1}{2}A_{\rm indef},
\end{align}
we indeed see that indefinite order leaves an imprint in linearly independent components of the joint system-control state of Eq.~\eqref{eq:SCstate_+input}.

Even though the global state $\rho_{SC}^{\mathcal{M}\leftrightarrow\mathcal{N}}$ may carry terms associated with SCO, the \textit{local} state of the central system $S$ is, up until this point, oblivious to such phenomenon. In fact, tracing out the control in Eq.~\eqref{eq:SCstate_+input} leads to $A_{++}+A_{--}=A_{\rm def}$. In order for SCO to manifest upon the state of $S$ locally, a post-selection of the control state must be performed. This later measurement, if implemented in the computational basis (associated with the operator $\sigma_z$), defines an order for the operations. Indeed, since the switch map associates each element of the control basis of $\sigma_z$ with a definite order of application of the maps $\mathcal{M}$ and $\mathcal{N}$, a notable property of the switch channel is that it maps $\sigma_z$-incoherent states of the control onto $\sigma_z$-incoherent states that can be associated with a classical mixture of orders. Because of this, it is common to consider a post-selection of the control in a state that has maximum coherence in the $\sigma_z$ basis, e.g., the eigenstates $\ket{+}$ or $\ket{-}$ of the $x$-Pauli operator. From Eq.~\eqref{eq:SCstate_+input}, the probability of each outcome in the post-selection is simply
\begin{align}
\label{eq:post_prob_UsualSwitch}
    p_{\rm post}(\pm)=\tr_{SC}\left\{\left(\mathds{1}_S \otimes |\pm\rangle \langle\pm|_C\right ) \rho_{SC}^{\mathcal{M} \leftrightarrow \mathcal{N}}\right\} = \tr_S\left\{A_{\pm\pm}\right\}.
\end{align}
Given that a post-selection of the control was made (and therefore $p_{\rm post}(\pm)>0$), the conditional state of the system $S$ is
\begin{align}
    \label{eq:conditionalS_+input}\rho_{S,\pm}^{\mathcal{M}\leftrightarrow\mathcal{N}}= \frac{A_{\pm\pm}}{\tr\left\{A_{\pm\pm}\right\}}.
\end{align}
Given the fact that $\rho_{S,\pm}^{\mathcal{M}\leftrightarrow\mathcal{N}}$ is proportional to $A_{\pm\pm}$, we know that these conditional states, obtained after the post-selection of the control, carry terms associated with SCO.

\section{Results}
\label{sec:results}

\subsection{Quantum switch with open control}

Consider now an interaction of the control with an environment $E$ right before its post-selection. For that, we will make use of the collisional model, in which the environment is represented as a stream of qubits in a well-defined Gibbs state, i.e., $\Theta_{E} = \text{exp}(-\beta_E H_{E})/Z_{E}$, where $H_{E}$ is the bare Hamiltonian of each environment qubit, $\beta_E=1/T_E$ is the inverse of temperature $T_E$, and $Z_{E} = \tr\{\text{exp}(-\beta_E H_{E})\}$ is the partition function. Both the Hamiltonians of $C$ and $E$ are assumed to be resonant, and the eigenbasis of $H_C$ is assumed to coincide with the post-selection basis, i.e., $\{\ket{+}_{C},\ket{-}_{C}\}$. Then, we can write $H_{C, E} = -\omega \sigma_x^{C, E}/2$ for a certain $\omega$. Observe that the choice of eigenbasis $\{\ket{+}_{E},\ket{-}_{E}\}$ for $H_{E}$ does not constitute a further restriction of our model since the reference frame for the environment can be chosen arbitrarily. Meanwhile, the specified control Hamiltonian guarantees that, up to a phase, the post-selection basis is invariant over its free dynamics. This model is represented in Fig.~\ref{fig:generalscheme}.

Between the controlled operation and the post-selection, the environment qubits interact one by one with the control qubit, i.e., each environment qubit couples with the control through some interaction Hamiltonian for a finite time $\tau$. After each interaction (also referred to as collision), the composite system-control state is updated according to
\begin{equation}
\label{eq:stateupdatecontrolcoll}
    \rho_{SC}^n = \tr_{E}\left \{ U \left (\rho_{SC}^{n-1} \otimes \rho_{E} \right ) U^\dagger \right \},
\end{equation}
where $n$ is the number of collisions, the trace is applied over the environment degrees of freedom, and $U = \text{exp}(-i\tau H_\text{tot})$ is the joint time evolution operator with
\begin{equation}\label{totalHamiltonian}
    H_\text{tot} = H_S + H_C + H_{E} + V_{CE}
\end{equation}
the total Hamiltonian, $H_\alpha$ the bare Hamiltonian of subsystem $\alpha$, and $V_{CE}$ the interaction between control and each environment qubit. The latter will be assumed to have the following form
\begin{equation}\label{interaction}
    V_{CE} = \frac{g}{2} \left( \sigma_z^C \sigma_z^{E} + \sigma_y^C\sigma_y^{E}  \right),
\end{equation}
where $g$ is the coupling strength, i.e., the interaction has an isotropic structure. In the collisional model, we assume that ${0<g\tau\ll1}$.

Note that Eq.~\eqref{interaction} can also be expressed as $g\left(|+\rangle \langle-|_{C}\otimes |-\rangle \langle +|_{E}+\text{h.c.}\right)$, with ``h.c.'' denoting Hermitian conjugate. This corresponds to the usual Jaynes-Cummings coupling for a reservoir of qubits~\cite{Ciccarello_2013} (in the ${\ket{\pm}}$ basis representation). Such form represents a standard system-reservoir coupling describing the exchange of excitation \cite{cusumano2022quantum, ciccarello2022quantum}. Moreover, this model is of thermodynamic interest since it commutes with the sum of the local bare Hamiltonians of the control and environment auxiliary systems, i.e., $[H_C + H_{E}, V_{CE}]_- = 0$. While the former assures excitation conservation, the latter satisfies the \textit{strict energy conservation} during the energy flow \cite{dann2021open}, which also implies that no work is performed during the collisions~\cite{Molitor2020}. In simpler terms, the mean energy of the interaction is constant, and all energy that leaves the control enters the environment ancilla, and vice versa.

Finally, starting with the post-QS state in Eq.~\eqref{eq:statepostQSSimonov} and considering the initial state ${\rho_{C}=|+\rangle\langle+|_{C}}$ for the control, the difference equation in Eq.~\eqref{eq:stateupdatecontrolcoll} can be solved, leading to the following composite system-control state after $n$ collisions:
\begin{equation}
\label{eq:GenExpAfterColl}
    \begin{aligned}
        \rho_{SC}^{n}= &\; \mathcal{B}_{++}(n)\otimes|+\rangle\langle+|_C+\mathcal{B}_{+-}(n)\otimes|+\rangle \langle-|_C\\
        &+\mathcal{B}_{-+}(n)\otimes|-\rangle \langle+|_C+\mathcal{B}_{--}(n)\otimes|-\rangle \langle-|_C,
    \end{aligned}
\end{equation}
where $\mathcal{B}_{+-}(n)=\mathcal{B}_{-+}^{\dagger}(n) \coloneqq e^{in\tau\omega}\cos^{n}(\text{g\ensuremath{\tau}})U_{S}^{n}A_{+-}U_{S}^{\dagger n}$ and
\begin{equation}
\label{eq:Bpmpm}
    \begin{split}
        \mathcal{B}_{\pm\pm}(n)\coloneqq &\; \frac{1}{2}\left\{1\pm f_{E}\left[1-\cos^{2n}(\text{g\ensuremath{\tau}})\right]\right\}A_{\text{def}}^{n}\\
        &\pm\frac{1}{2}\cos^{2n}(\text{g\ensuremath{\tau}})A_{\text{indef}}^{n}
    \end{split}
\end{equation}
with $U_{S}\coloneqq\text{exp}(-i \tau H_S)$ being the time-evolution operator of the system, $A_{\text{def}}^{n}\coloneqq U_{S}^{n}A_{\textrm{def}}U_{S}^{\dagger n}$,  $A_{\text{indef}}^{n}\coloneqq U_{S}^{n}A_{\textrm{indef}}U_{S}^{\dagger n}$, and $f_E \coloneqq \tanh (\beta_E \omega/2)$. Note that $A_{\rm def}^n$ and $A_{\rm indef}^n$ are the time-evolved versions of $A_{\rm def}$ and $A_{\rm indef}$ for a period $\tau$, respectively.
Eq.~\eqref{eq:GenExpAfterColl} constitutes our main result, as it gives the joint system-control state after $n$ collisions, for any implementation of the 2-quantum switch with its control affected by the environment. Hence, we can now examine how this open control affects the operation of the QS.

Note that Eq.~\eqref{eq:GenExpAfterColl} is analogous to Eq.~\eqref{eq:SCstate_+input}, differing only by the change $A_{xy}\to \mathcal{B}_{xy}(n)$. Given that this is the only formal change on $\rho_{SC}^{\mathcal{M}\leftrightarrow\mathcal{N}}$ due to the open control, we can understand the effect of the environment by analyzing how these coefficients $\mathcal{B}_{xy}$ behave and how they compare to $A_{xy}$. Before we do so, first note that
\begin{align}
\label{eq:Bxy=Axy}
    \mathcal{B}_{xy}(0)=A_{xy},
\end{align}
so the limit with the traditional closed control in Eq.~\eqref{eq:SCstate_+input} is reproduced when there are no collisions. Second, the local state of the system after $n$ collisions is given by
\begin{align}
\label{eq:asymptoticStateSC}
\rho_{S}^{n}&= \tr_C\left\{\rho_{SC}^n\right\}=\mathcal{B}_{++}(n)+\mathcal{B}_{--}(n)=A_{\text{def}}^{n}
\nonumber\\
&=\frac{1}{2}U_{S}^{n}\left[\mathcal{N}\circ\mathcal{M}(\rho_{S})+\mathcal{M}\circ\mathcal{N}(\rho_{S})\right]U_{S}^{\dagger n}
\end{align}
for \textit{all} $n$, where the last equality comes from Eq.~\eqref{eq:A_def} and the definition of $A_{\rm def}^n$. This shows that $\rho_S^n$ corresponds to the mixture of causally ordered quantum maps, unitarily evolved according to the local Hamiltonian, as expected. Third, in the limit of many collisions, $n\to\infty$, the joint system-control state is
\begin{align}
\label{eq: assymptoticSCstate}
     \lim_{n\to\infty}\rho_{SC}^n &= \left[\lim_{n\to\infty}\mathcal{B}_{++}(n)\right]\otimes|+\rangle \langle+|_C + \left[\lim_{n\to\infty}\mathcal{B}_{--}(n)\right]\otimes |-\rangle \langle-|_C\nonumber\\
     &= \left(\lim_{n\to\infty}A^n_{\rm def}\right)\otimes\left(\frac{1+f_E}{2}|+\rangle \langle+|_C + \frac{1-f_E}{2}|-\rangle \langle-|_C\right)\nonumber\\
     &=\rho^\infty_S\otimes\Theta_{\beta_{E}},
\end{align}
where we used that $\cos{\left(g\tau\right)}<1$ (recall that $0<g\tau\ll1$) to calculate the limits of $\mathcal{B}_{xy}(n)$, and  $\Theta_{\beta_{E}}$ corresponds to the thermal state of the control relative to the inverse of temperature $\beta_{E}$. That is, the joint state becomes a product one, showing that the correlations are suppressed in the asymptotic limit, with the control system being in the thermal state, as expected.

\subsection{Post-selection}

Since the local state of the system $S$ can only carry effects of SCO with a post-selection of the control, we now focus on the effect of the environment on such post-selection (again, in the $\ket{\pm}_C$ basis). The equations that give us the post-selection probabilities and conditional states can be obtained directly from Eq.~\eqref{eq:GenExpAfterColl} (and by direct analogy with Eqs.~\eqref{eq:post_prob_UsualSwitch} and ~\eqref{eq:conditionalS_+input}), i.e.,
\begin{align}
\label{eq:postselectionprobBpmpm}
    p^n_{\rm post}(\pm)=\tr\left\{\left(\mathds{1}_S \otimes |\pm\rangle \langle\pm|_C\right ) \rho_{SC}^{n}\right\} = \tr\left\{\mathcal{B}_{\pm\pm}(n)\right\},
\end{align}
and
\begin{align}
        \label{eq:conditionalStateBpmpm}
        \rho_{S,\pm}^{n}= \left[p^n_{\rm post}(\pm)\right]^{-1} \mathcal{B}_{\pm\pm}(n),
\end{align}
respectively.

\begin{figure}
    \center
    \includegraphics[width=\columnwidth]{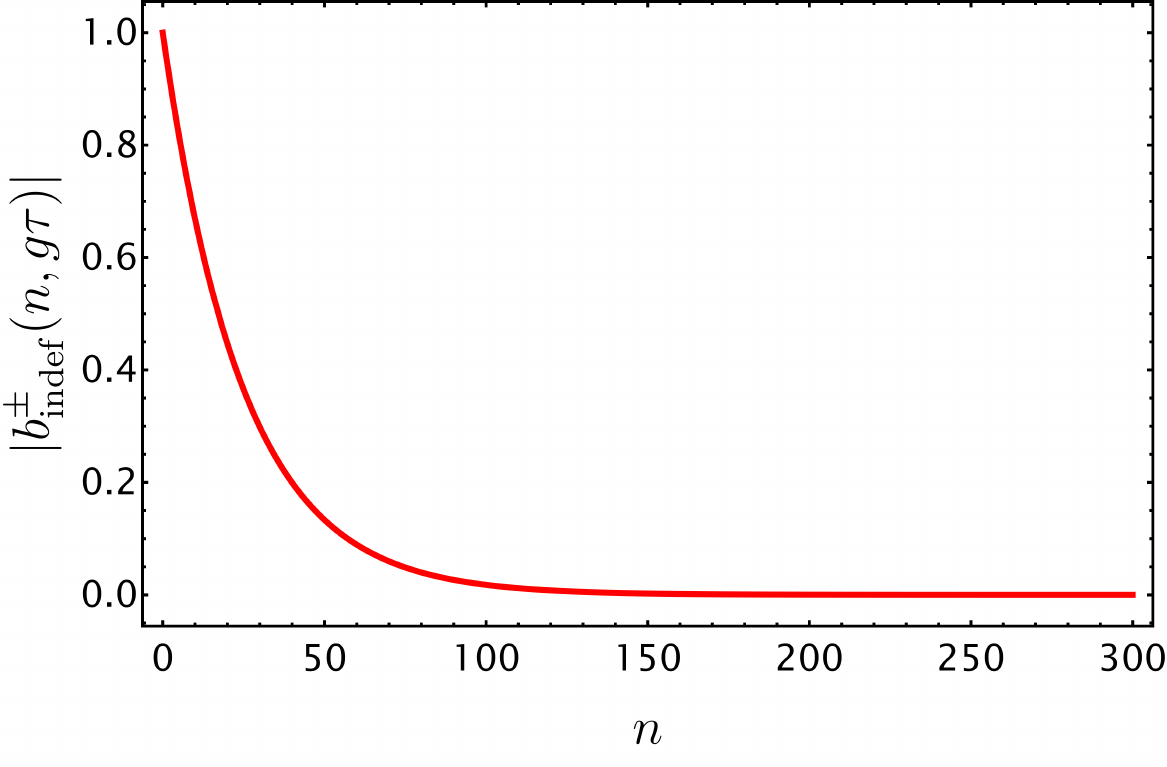}
    \caption{ \textbf{Monotonic behavior of SCO features as a function of $n$}. It is assumed that $g\tau=0.2$. As noted in the text, the SCO features encoded in $|b_{\rm indef}^\pm(n,g\tau)|$ decrease monotonically with increasing number of collisions $n$, independently of $T_E$ and $\omega$. Note that, for the model introduced here, the behavior of the SCO effects generated by two arbitrary quantum operations does not depend on the environment temperature.}
    \label{fig: PlotMonitoramentoMINUS}
\end{figure}

We note that Eqs~\eqref{eq:postselectionprobBpmpm} and \eqref{eq:conditionalStateBpmpm} make explicit what was already anticipated: In order to understand how SCO is affected by the environment and the different parameters of this interaction (such as the number of collisions, temperature, etc), we only need to analyze how these parameters change the operators $\mathcal{B}_{\pm\pm}$. To do so, we first rewrite these as
\begin{align}
\label{eq:BpmpmIntermsofbdefindef}
    \mathcal{B}_{\pm\pm}(n)=\frac{b^{\pm}_{\rm def}(n,f_E,g\tau)}{2}A_{\rm def}^n + \frac{b^{\pm}_{\rm indef}(n,g\tau)}{2}A^n_{\rm indef},
\end{align}
where, by Eq.~\eqref{eq:Bpmpm}, 
\begin{equation}
    \label{eq:bdef&indef}
    \begin{aligned}
        b^{\pm}_{\rm def}(n,f_E,g\tau) &:= 1\pm f_{E}\left[1-\cos^{2n}(\text{g\ensuremath{\tau}})\right],\\
        b^{\pm}_{\rm indef}(n,g\tau) &:=\pm\cos^{2n}(\text{g\ensuremath{\tau}}).
    \end{aligned}
\end{equation}
Eq.~\eqref{eq:BpmpmIntermsofbdefindef} shows that $b_{\rm indef}^\pm$ modulates the impact of SCO in the post-selection: The higher $|b^{\pm}_{\rm indef}|$, the higher the SCO effect of a given QS; whenever this term is null, there is no SCO effect at all. In fact, we can get valuable information from the dependence of $b_{\rm indef}^\pm$ on $n$: \emph{Collisions monotonically decrease the effect of SCO}. Indeed, since $0<{g\tau}\ll 1$,
\begin{align}
    |b^\pm_{\rm indef}(n+1,g\tau)|<|b^\pm_{\rm indef}(n,g\tau)|.
\end{align}
Moreover, the maximum value of $|b^\pm_{\rm indef}|$, obtained at $n=0$, is $|b^{\pm}_{\rm indef}(n=0,g\tau)|=1$, while in the asymptotic limit we get $b^{\pm}_{\rm indef}(n\to\infty,g\tau)=0$, in accordance with Eq.~\eqref{eq:asymptoticStateSC}. This qualitative behaviour of $|b_{\rm indef}^\pm(n,g\tau)|$ is depicted in Fig~\ref{fig: PlotMonitoramentoMINUS}. Surprisingly, neither the temperature $T_{E}$ of the bath nor the frequency $\omega$ impacts the presence or absence of SCO effects. Indeed, $b_{\rm indef}^\pm(n,g\tau)$ has no dependence on these parameters whatsoever.

Such independence, however, does not imply that these quantities have no effect on the post-selection as a whole. Indeed, $b^\pm_{\rm def}$ is affected by these parameters through $f_E$, which, in turn, impacts $\mathcal{B}_{\pm\pm}$, reflecting on the post-selection probability $p^n_{\rm post}(\pm)$ and the conditional states $\rho_{S,\pm}^n$. Using the fact that $\tr\left\{A_{\rm indef}^n\right\}=\tr\left\{A_{\rm indef}\right\}$ and $\tr\left\{A_{\rm def}^n\right\}=1$, Eq.~\eqref{eq:postselectionprobBpmpm} can be rewritten as
\begin{equation}
     \label{eqs:postselectionbdebindef}
    p^n_{\rm post}(\pm)= \frac{1}{2} \left[b^\pm_{\rm def}(n,f_E,g\tau) + b^\pm_{\rm indef}(n,g\tau)\tr\left\{A_{\rm indef}\right\}\right].
\end{equation}
Also, with Eqs.~\eqref{eq:conditionalStateBpmpm} and~\eqref{eq:BpmpmIntermsofbdefindef}, we have
\begin{equation}
     \label{eq:conditionalstatebdefbindef}
    \rho_{S,\pm}^n = \left[p^n_{\rm post}(\pm)\right]^{-1} \left[\frac{b^{\pm}_{\rm def}(n,f_E,g\tau)}{2}A_{\rm def}^n + \frac{b^{\pm}_{\rm indef}(n,g\tau)}{2}A^n_{\rm indef}\right].
\end{equation}

Let us look into the impact of the two extreme temperature regimes on these objects. The most interesting case is the condition $T_E\to 0$ (i.e., $\beta_E\to\infty$ and $f_E\to 1$), particularly for the post-selection on $\ket{-}_C$. In this case, it can be seen that $b^{-}_{\rm def}=-b^{-}_{\rm indef}$. Then, from Eq.~\eqref{eq:conditionalstatebdefbindef}, we get
\begin{align}
    \lim_{\beta_E\to\infty}\rho^n_{S,-}=\frac{b^{-}_{\rm def}\left(A_{\rm def}-A_{\rm indef}\right)}{b^{-}_{\rm def}
    \left(1-\tr\left\{A_{\rm indef}\right\}\right)}=\frac{A^n_{--}}{\tr\left\{A_{--}^n\right\}}.
    \label{eq:sistemS-post}
\end{align}
Therefore, independently of the particular channels inside of the QS, the conditional state $\rho^n_{S,-}$ \emph{is completely shielded from the environmental interactions}. After a post-selection on $\ket{-}_C$, the obtained state of $S$ is oblivious to the impact of the environment on the control, being the result of local evolution independently of the number of collisions. This limit must be considered with care, though, since the probability of attaining such a post-selection goes to zero as $n$ increases since
\begin{eqnarray}
    \lim_{\beta_E\to\infty }p^n_{\rm post}(-) &=& \cos^{2n}{\left(g\tau\right)}\frac{\left(1-\tr\left\{A_{\rm indef}\right\}\right)}{2}\\
    &=&  \cos^{2n}{\left(g\tau\right)}p^0_{\rm post}(-), 
\end{eqnarray}
where $p^0_{\rm post}(-)$ is the probability of post-selection without any environmental action on the control\footnote{Intuitively, in the limit $T_E\to 0$, the bath is initialized in $|+\rangle \langle +|_E$, which induces thermalization of the control in the state $|+\rangle\langle+|_C$, reducing the probability of post-selection on $\ket{-}_C$.}.

The low-temperature regime does not provide such a shielding effect in the case of post-selection in the $\ket{+}_C$ outcome, which occurs with increasing probability as $n$ grows. Indeed, in that case $b^+_{\rm def}=2 - b^+_{\rm indef}$, and substituting this into Eq.~\eqref{eq:conditionalstatebdefbindef} shows that there is always an impact of the environmental interactions on the state $\rho^n_{S,+}$. Specifically, as we already know, the SCO contribution is suppressed with increasing $n$, and only definite order terms survive.

\begin{figure}
    \center
    \includegraphics[width=\columnwidth]{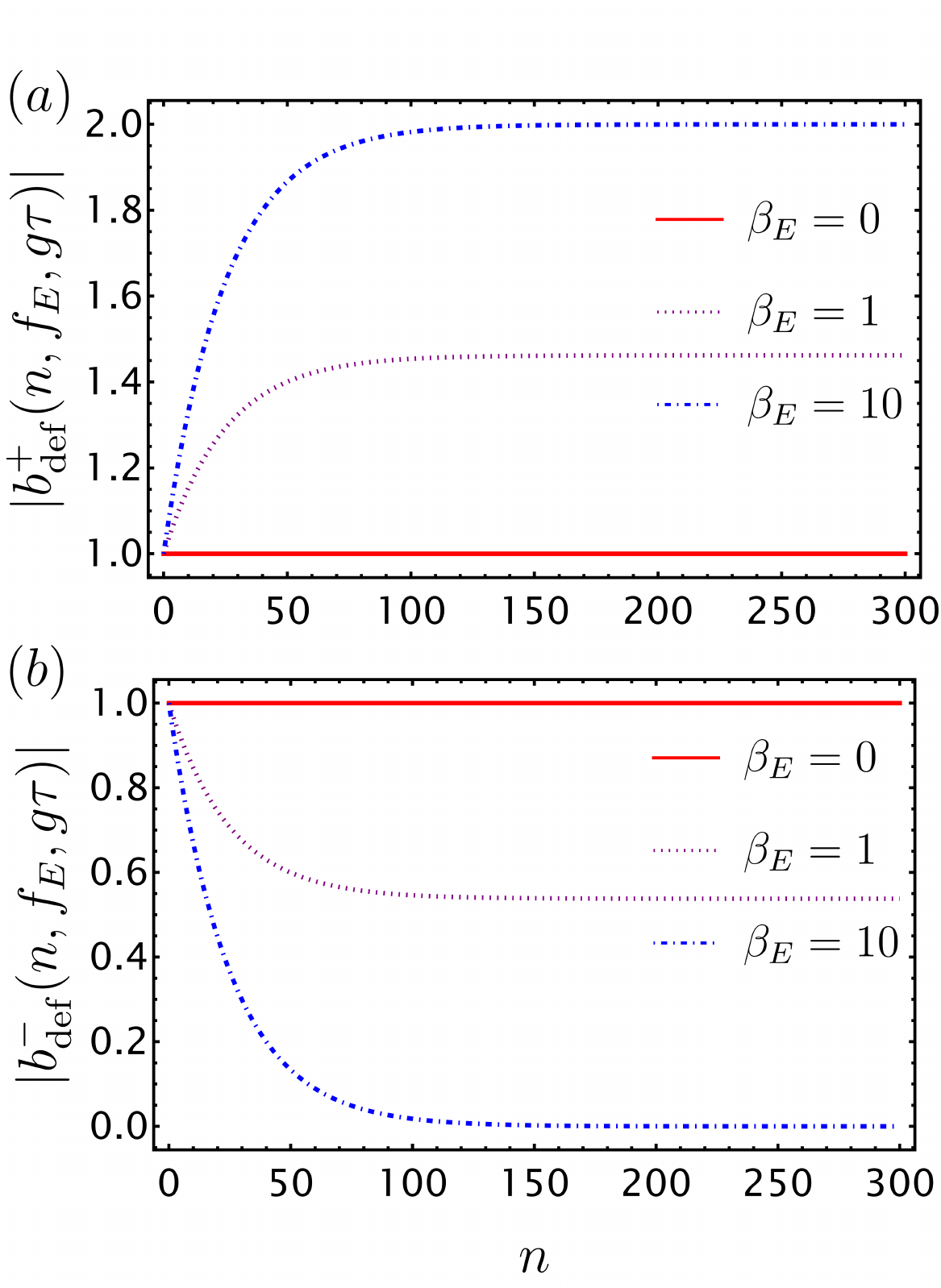}
    \caption{\textbf{Thermal effects for definite order features as a function of $n$}. It is assumed $\omega=1$, and $g\tau=0.2$. $\beta_E$ has units with the inverse of $\omega$. The red solid curve represents $\beta_E=0$, the dotted purple curve represents $\beta_E=1$, and the dot-dashed blue curve represents $\beta_E=10$. In low-temperature regimes, the definite order terms for the $\ket{-}$-output are suppressed compared to the $\ket{+}$-output. In the limit $T_E\to \infty$, both quantities converge to the same value.}
    \label{PlotMonitoramentoDef}
\end{figure}

Finally, in the $T_E\to \infty$ case (i.e., $\beta_E\to 0$ and $f_E\to 0$), $b^{\pm}_{\rm def}=1$ for all values of $n$, implying that the definite order terms are not affected by the interactions. The latter only suppresses the SCO contributions to $\rho^n_{S,\pm}$, thus providing no shielding effect. As another difference to the $T_{E}\to 0$ regime, in which the probability of the post-selection in the $\ket{-}_C$ state always decreases with $n$, in the high-temperature regime the collisions may affect the post-selection probabilities in different ways depending on the model. That is, whether $p_{\rm post}^n(+)$ or $p_{\rm post}^n(-)$ increases with $n$ depends on the particular implementation of the QS, as it depends on the sign of $\tr\left\{A_{\rm indef}\right\}$, as evidenced by
\begin{align}
\lim_{\beta_E\to 0}p^n_{\rm post}(\pm) = \frac{1+b^{\pm}_{\rm indef}\tr\left\{A_{\rm indef}\right\}}{2}.
\end{align}
In Fig.~\ref{PlotMonitoramentoDef}, we show some examples of the behavior of $b^{\pm}_{\rm def}$ with temperature, where an intermediate and the limit cases described above can be seen.

In this section, we have analyzed the general, model-independent results coming from Eq.~\eqref{eq:GenExpAfterColl}. However, the expression for the conditional states and actual value of $p_{\rm post}^n$ depend on the particular implementation of the QS. These results are organized in Table~\ref{table:TemperatureQS}. To illustrate our results, we present two particular examples in the Appendices, namely, a QS of MUB-monitoring channels and the QS-driven refrigerator present in Ref.~\cite{Felce2020}.

\begin{table}
\resizebox{\columnwidth}{!}{%
\begin{tabular}{@{}lcccc@{}}
\cmidrule(l){2-5}
\multicolumn{1}{c}{}                                                          & \multicolumn{4}{c}{\textbf{Impact of Temperature on QS as $n\to\infty$}}                                                                                                                                     \\ \cmidrule(l){2-5} 
                                                                              & $\rho^n_{S,-}$                                                                                 & $p^n_{\rm post}(-)$                        & $\rho^n_{S,+}$                         & $p^n_{\rm post}(+)$   \\ \midrule
\textbf{\begin{tabular}[c]{@{}l@{}}Low-$T_{E}$\\ ($\beta_E\to\infty$)\end{tabular}} & \multicolumn{1}{c|}{\begin{tabular}[c]{@{}c@{}}Shielding effect: \\ SCO survives\end{tabular}} & \multicolumn{1}{c|}{Goes to $0$}           & \multicolumn{1}{c|}{SCO is suppressed} & Goes to $1$           \\ \midrule
\textbf{\begin{tabular}[c]{@{}l@{}}High-$T_{E}$\\ ($\beta_E\to 0$)\end{tabular}}    & \multicolumn{1}{c|}{SCO is suppressed}                                                         & \multicolumn{1}{c|}{Goes to $\frac{1}{2}$} & \multicolumn{1}{c|}{SCO is suppressed} & Goes to $\frac{1}{2}$ \\ \bottomrule
\end{tabular}
}
\caption{\textbf{Limiting cases for the open control model.} For high external temperatures $T_E$, the SCO contribution for both post-selected states is suppressed with probabilities approaching $1/2$. For low temperatures, while the SCO contribution for $\rho^n_{S,+}$ is suppressed, the contribution for $\rho^n_{S,-}$ shows to be shielded against the environmental interactions, i.e., independent of the number of collisions. Despite the shielding, the probability of post-selecting $\ket{-}_C$ and $\ket{+}_C$ goes to $0$ and $1$, respectively.}
\label{table:TemperatureQS}
\end{table}

\section{Discussion}
\label{sec:discussion}

In this work, we have characterized the environmental-induced instabilities mediated by the control system in the QS of two arbitrary quantum operations. Having an open control (with the Jaynes--Cummings-like interaction presented here) always negatively impacts the contribution of SCO in a QS. Perhaps surprisingly, the presence of SCO contributions is \emph{not} affected by the temperature, independently of the particular QS being implemented. 
However, the bath properties and coupling specifics ($T_{E}$, $\omega$, and $g\tau$) \emph{do} affect the post-selection probabilities and conditional states.
In the low-temperature case, the bath induces an interesting asymmetry, where $\rho^n_{S,-}$ is shielded from the impact of the environment---even though such a post-selection becomes more unlikely with each additional collision. However, the favored outcome in the low-temperature regime, $\ket{+}_C$, is always affected by the collisions, which suppress SCO. Both outcomes are similarly affected in the high-temperature regime, with SCO being suppressed and the post-selection probability for each outcome becoming equally likely in the limit $n\to \infty$, as summarized in Table~\ref{table:TemperatureQS}. Therefore, in any implementation of the QS where the control may be subject to environmental interactions, one should expect a qualitatively different behavior.

From a thermodynamic perspective, the environment auxiliary systems play major roles, i.e., they both exchange an energy amount $\Delta U^{n}_E$ with the control and induce entropic changes. Such couplings are the root of the process of dissipation and irreversibility undergone by the control. As previously mentioned, given the Hamiltonian structure of Eq.~\eqref{totalHamiltonian}, energy conservation holds during each collision in a way that no work is performed and $\Delta U^{n}_C\coloneqq\tr\left\{ \left ( \rho_C^n - \rho_C^{n-1} \right ) H_C \right \} \equiv -\Delta U^{n}_E$. Hence, the total heat transferred to the control after $n$ collisions can be cast as
\begin{equation}
    \frac{2}{\omega}\mathcal{Q}_{CE}^{n}=\left(b_{{\rm def}}^{-}(n,f_{E},g\tau)-1\right)+\left(b_{{\rm indef}}^{-}(n,g\tau)+1\right)\tr\left\{ A_{{\rm indef}}\right\} .
\end{equation}
Therefore, the \textit{entropy production} of the composite $SC$ state is given by the difference between the entropy change $\Delta \mathcal{S}_{SC}^n = \mathcal{S}(\rho_{SC}^n) - \mathcal{S}(\rho_{SC}^{0})$ and the entropy flux $\beta_E \mathcal{Q}_{CE}^n$ accompanied by the heat~\cite{landi2021irreversible}, i.e.,
\begin{equation}
    \Sigma_{SC}^n = \Delta \mathcal{S}_{SC}^n - \beta_E \mathcal{Q}_{CE}^n,
\end{equation}
where $\mathcal{S}(\rho) \coloneqq -\tr\{\rho \ln \rho\}$ is the von Neumann entropy of $\rho$. This quantifies in thermodynamic terms the irreversibility of the open system dynamics of the control with the environment. 

The main result of this work is to provide a methodology for considering an open control quantum switch that can be employed to analyze the effect of the environmental instabilities in the figure of merits of several quantum switch-based protocols. To illustrate this, in the Appendices, we employ our framework to analyze the consequences of having an open control in two distinct contexts, namely, the QS of monitoring of mutually unbiased bases (MUBs) and a quantum refrigerator induced by SCO~\cite{felce2020quantum}. In particular, we consider
the \textit{available information} after $n$ collisions and post-selection of the control
\begin{equation}
\label{eq:infosystempostmeas}
    \mathcal{I}(\rho_{S,\pm}^n) = \ln d_S - \mathcal{S}(\rho_{S,\pm}^n),
\end{equation}
where $d_S$ is the dimension of the Hilbert space of the system $\mathcal{H}_S$. This quantity is shown to have asymmetric behaviors according to the post-selection, i.e., while a post-selection in $\ket{+}_C$ preserves the monotonic decreasing relation with the measurement strength of each map, the $\ket{-}_C$ post-selection breaks such monotonicity.

The significant instabilities identified in this study, particularly in the asymmetric input-output configuration of the control, hold special relevance for protocols relying solely on this post-selection as the refrigerator induced by SCO~\cite{felce2020quantum}. In this model, detailed in the Appendices, we demonstrate how these instabilities consistently degrade its performance. An intriguing avenue for future research would involve developing protocols resilient to such instabilities or considering whether they can be identified as an additional resource. Moreover, our collisional model can also be adjusted to incorporate features such as non-thermal baths with quantum coherence~\cite{Rodrigues2019} and non-Markovian interactions~\cite{Camasca2021}, for instance.

\begin{appendix}

\section{Monitoring of mutually unbiased bases}

Monitoring maps are CPTP maps that interpolate between weak and strong non-selective measurements. They can be defined as~\cite{oreshkov2005weak, dieguez2018information}
\begin{equation}
    \mathcal{M}_{\mathcal{O}}^{\epsilon}(\rho_S) \coloneqq (1-\epsilon) \rho_S + \epsilon\, \Phi_{\mathcal{O}}(\rho_S),
    \label{eq:mon-map}
\end{equation}
where $0\leqslant\epsilon\leqslant1$ is the measurement strength and the map $\Phi_{\mathcal{O}}$ is a dephasing of system $S$ in the eigenbasis of the operator $\mathcal{O}=\sum_\alpha \alpha \, \mathcal{O}_\alpha$, i.e.,
\begin{equation}
    \label{eq:reality}
    \Phi_\mathcal{O}(\rho_S) \coloneqq \sum_{\alpha} \mathcal{O}_\alpha  \rho_S  \mathcal{O}_\alpha =\sum_\alpha p_\alpha \mathcal{O}_\alpha
\end{equation}
with $p_\alpha = \tr\{\mathcal{O}_\alpha \rho_S\}$, and $\mathcal{O}_\alpha$ are projectors such that $\mathcal{O}_\alpha \mathcal{O}_{\alpha'}=\delta_{\alpha \alpha'}\mathcal{O}_\alpha$. Map $\Phi_O$ can be interpreted as the projective measurement of observable $O$ without having its outcome revealed. A possible choice of Kraus decomposition for this operation is $K_0 = \sqrt{1-\epsilon} \,\mathds{1}_{S}$ and $K_j = \sqrt{\epsilon}\, \mathcal{O}_j$. These maps satisfy the property
\begin{equation}
    \mathcal{M}^{\epsilon}_{\mathcal{O}} \circ \mathcal{M}^{\epsilon'}_{\mathcal{O}}(\rho) = \mathcal{M}^{\epsilon''}_{\mathcal{O}}(\rho),
    \label{eq:monitoring-same-O}
\end{equation}
where $\epsilon''=\epsilon+\epsilon'-\epsilon\epsilon'$ \cite{dieguez2018information}.

Before discussing the application of these channels in SCO, it is worth noting that the above operations always decrease the amount of information in the reference frame of the system. By the concavity of the von Neumann entropy, it is easy to see that $\mathcal{S}(\mathcal{M}^{\epsilon}_{\mathcal{O}} (\rho_S))\geqslant (1-\epsilon)\mathcal{S}(\rho_S)+\epsilon \mathcal{S}(\Phi_{\mathcal{O}}(\rho_S))$ and
\begin{equation}
   \mathcal{I}(\rho_S)-\mathcal{I}(\mathcal{M}^{\epsilon}_{\mathcal{O}}(\rho_S)) \geqslant \epsilon  \mathfrak{C}_{\mathcal{O}}(\rho_S)\geqslant 0,
\end{equation}
where $\mathfrak{C}_{\mathcal{O}}(\rho) \coloneqq \mathcal{S}(\Phi_{\mathcal{O}}(\rho_S))-\mathcal{S}(\rho_S)$ is the relative entropy of coherence related with the observable $\mathcal{O}$. Therefore, the available information exhibits a monotonic relation as a function of the monitoring strength~\cite{dieguez2018information}.

Here, we are interested in the scenario in which the operators $\mathcal{O}$ and $\mathcal{O}'$ are associated with MUBs.
In this case, it can be verified that $\Phi_{\mathcal{O}}\circ\Phi_{\mathcal{O}'}(\rho_S) = \Phi_{\mathcal{O}'}\circ\Phi_{\mathcal{O}}(\rho_S) = \mathds{1}_S/d_S$.
This implies that two consecutive monitorings of MUBs commute, in the sense that $\mathcal{M}^{\epsilon'}_{\mathcal{O}'} \circ \mathcal{M}^{\epsilon}_{\mathcal{O}}(\rho_S)=\mathcal{M}^{\epsilon}_{\mathcal{O}} \circ\mathcal{M}^{\epsilon'}_{\mathcal{O}'}(\rho_S)$ for every measurement strengths $\epsilon$ and $\epsilon'$. The output state reads
\begin{equation}
\begin{aligned}
   \mathcal{M}^{\epsilon'}_{\mathcal{O}'} \circ \mathcal{M}^{\epsilon}_{\mathcal{O}}(\rho_S)= &\; (1-\epsilon)(1-\epsilon')\rho_S +\epsilon(1-\epsilon')\Phi_{\mathcal{O}}(\rho_S)\\
    &+ \epsilon'(1-\epsilon)\Phi_{\mathcal{O}'}(\rho_S)+\epsilon\epsilon' \mathds{1}_S/d_S.
       \label{eq:monitoring-mubs}
\end{aligned}
\end{equation}
Employing the concavity of the von Neumann entropy it follows that
\begin{equation}
\begin{aligned}
  \mathcal{I}(\rho_S)- \mathcal{I}(\mathcal{M}^{\epsilon'}_{\mathcal{O}'} \circ \mathcal{M}^{\epsilon}_{\mathcal{O}}(\rho_S)) \geqslant &\; \epsilon\epsilon' \mathcal{I}(\rho_S) +\epsilon(1-\epsilon')\mathfrak{C}_{\mathcal{O}}(\rho_S)\\
    &+ \epsilon'(1-\epsilon)\mathfrak{C}_{\mathcal{O'}}(\rho_S).
\end{aligned}
\end{equation}
which is a non-negative quantity from the positivity of the available information, the relative coherence for each basis, and $0\leqslant \epsilon,\epsilon'\leqslant 1$. Therefore, we conclude that
\begin{equation}
    \mathcal{I}(\rho_S)\geqslant \mathcal{I}(\mathcal{M}^{\epsilon'}_{\mathcal{O}'} \circ \mathcal{M}^{\epsilon}_{\mathcal{O}}(\rho_S)),
\end{equation}
which implies the monotonicity of the available information under consecutive monitoring.

For simplicity, from now on we set $\epsilon=\epsilon'$ to discuss the quantum switch of these monitoring maps. Consider the general equation after QS and collisions of control with an environment at inverse temperature $\beta$ in Eq. \eqref{eq:GenExpAfterColl} with $\mathcal{M} = \mathcal{M}_{\mathcal{O}}^{\epsilon}$ and $\mathcal{N} = \mathcal{M}_{\mathcal{O}'}^{\epsilon}$ (same monitoring strength $\epsilon = \epsilon'$), where the operators are $\mathcal{O}=\sum_{\alpha} \alpha_i \mathcal{O}_i$ and $\mathcal{O}'=\sum_{i} \alpha_i' \mathcal{O}'_{i}$ ($\mathcal{O}_i = |o_i\rangle \langle o_i|$ and $\mathcal{O'}_i = |o_i'\rangle \langle o_i'|$ are projectors onto the bases $\{\ket{o_i}\}_i$ and $\{\ket{o'_i}\}_i$, respectively). In this case, $M_0 = \sqrt{1-\epsilon}\, \mathds{1}_{S}$, $M_j = \sqrt{\epsilon}\, \mathcal{O}_j$, $N_0 = \sqrt{1-\epsilon}\, \mathds{1}_{S}$, and $N_j = \sqrt{\epsilon}\, \mathcal{O}'_j$.

A trivial case is when $\mathcal{O}' = \mathcal{O}$. Because of the property in Eq. \eqref{eq:monitoring-same-O}, it can be checked that the switch map reduces to $\mathcal{M}^{\epsilon'}_{\mathcal{O}} \otimes \mathds{1}_C$, where $\epsilon'=2\epsilon - \epsilon^2$. This conclusion and the property in Eq. \eqref{eq:monitoring-mubs} may lead someone to wrongly believe that a similar result holds when the eigenbases of $\mathcal{O}$ and $\mathcal{O}'$ are MUBs, for which $\langle o_i | o_j'\rangle  = e^{i \phi_{ij}}/\sqrt{d_S}$. However, this is not the case. Indeed, from Eqs.~\eqref{eq:A_def}, \eqref{eq: A_indef}, and~\eqref{eq:sistemS-post}, we have
\begin{equation}
    \begin{aligned}
        \rho_{S,\pm}^{n}=&\;\frac{b_{{\rm def}}^{\pm}(n,f_{E},g\tau)+b_{{\rm indef}}^{\pm}(n,g\tau)}{2p_{{\rm post}}^{n}(\pm)}U_{S}^{n}\mathcal{M}_{\mathcal{O}'}^{\epsilon}\circ\mathcal{M}_{\mathcal{O}}^{\epsilon}(\rho_{S})U_{S}^{\dagger n}\\&+\epsilon^{2}\frac{b_{{\rm indef}}^{\pm}(n,g\tau)}{2d_{S}p_{{\rm post}}^{n}(\pm)}\\&\cross\left[\frac{1}{2}\sum_{i,j}U_{S}^{n}\left(e^{2i\phi_{ij}}|o_{i}\rangle\langle o_{j}'|\rho_{S}|o_{i}\rangle\langle o_{j}'|+\text{h.c.}\right)U_{S}^{\dagger n}-\mathds{1}_{S}\right],
    \end{aligned}
\end{equation}
where
\begin{equation}\label{eq:normalizationplusminus}
    p_{{\rm post}}^{n}(\pm)=\tr\left\{ \mathcal{B}_{\pm\pm}(n)\right\} =\frac{b_{{\rm def}}^{\pm}(n,f_{E},g\tau)}{2}+\frac{b_{{\rm indef}}^{\pm}(n,g\tau)}{2}\textrm{Re}\left\{ \chi\right\} 
\end{equation}
with $\textrm{Re}\left\{ \chi\right\} =\tr\left\{ A_{\textrm{indef}}\right\}$ and
\begin{equation}
    \chi = (1-\epsilon)^2 + 2\epsilon (1-\epsilon) + \frac{\epsilon^2}{d_S^{3/2}} \sum_{i,j} e^{i \phi_{ij}} \bra{o_j'}\rho_S\ket{o_i}.
\end{equation}
This result is valid for any finite-dimensional system. For simplicity, in the specific case that the system is a qubit in the initial state $\rho_S = |+\rangle \langle +|_S$, the Hamiltonian is in the $\sigma_x$ basis
and the observables are $\mathcal{O}=\sigma_z$ and $\mathcal{O}'=\sigma_x$, the state post-QS, collisions and post-selection reads
\begin{equation}
\begin{aligned}
        \rho_{S,\pm}^{n}=&\;\frac{1}{2p_{{\rm post}}^{n}(\pm)}\Bigg[\left(1-\frac{\epsilon}{2}\right)\left(b_{{\rm def}}^{\pm}(n,f_{E},g\tau)+b_{{\rm indef}}^{\pm}(n,g\tau)\right)|+\rangle\langle+|_{S}\\
        &+\frac{\epsilon}{2}\Bigg(b_{{\rm def}}^{\pm}(n,f_{E},g\tau)+b_{{\rm indef}}^{\pm}(n,g\tau)\left(1-\epsilon\right)\Bigg)|-\rangle\langle-|_{S}\Bigg]
\end{aligned}
\end{equation}
with $p_{{\rm post}}^{n}(\pm)=\frac{1}{2}\left[b_{{\rm def}}^{\pm}(n,f_{E},g\tau)+b_{{\rm indef}}^{\pm}(n,g\tau)\left(1-\frac{\epsilon^{2}}{2}\right)\right]$. One then sees that the post-selected state is diagonal in the eigenbasis of $\sigma_x$. Finally, we calculate the available information to analyze how much information is stored in this state as the number of collisions increases for different measurement strengths. Using Eq.~\eqref{eq:infosystempostmeas} we get
\begin{widetext}
\begin{equation}
\begin{aligned}
    \mathcal{I}(\rho_{S,\pm}^{n})=&\;\ln2-\frac{1}{2p_{{\rm post}}^{n}(\pm)}\Bigg[\left(\frac{\epsilon}{2}-1\right)\left(b_{{\rm def}}^{\pm}(n,f_{E},g\tau)+b_{{\rm indef}}^{\pm}(n,g\tau)\right)\ln\left(\frac{\left(1-\frac{\epsilon}{2}\right)\left(b_{{\rm def}}^{\pm}(n,f_{E},g\tau)+b_{{\rm indef}}^{\pm}(n,g\tau)\right)}{2p_{{\rm post}}^{n}(\pm)}\right)\\&-\frac{\epsilon}{2}\Bigg(b_{{\rm def}}^{\pm}(n,f_{E},g\tau)+b_{{\rm indef}}^{\pm}(n,g\tau)\left(1-\epsilon\right)\Bigg)\ln\left(\frac{\epsilon}{2}\frac{b_{{\rm def}}^{\pm}(n,f_{E},g\tau)+b_{{\rm indef}}^{\pm}(n,g\tau)\left(1-\epsilon\right)}{2p_{{\rm post}}^{n}(\pm)}\right)\Bigg]
\end{aligned}
\end{equation}
\end{widetext}
whose direct interpretation is not straightforward, and with the help of plots we shall analyze it. 

\begin{figure}
\centering
\includegraphics[width=\columnwidth]{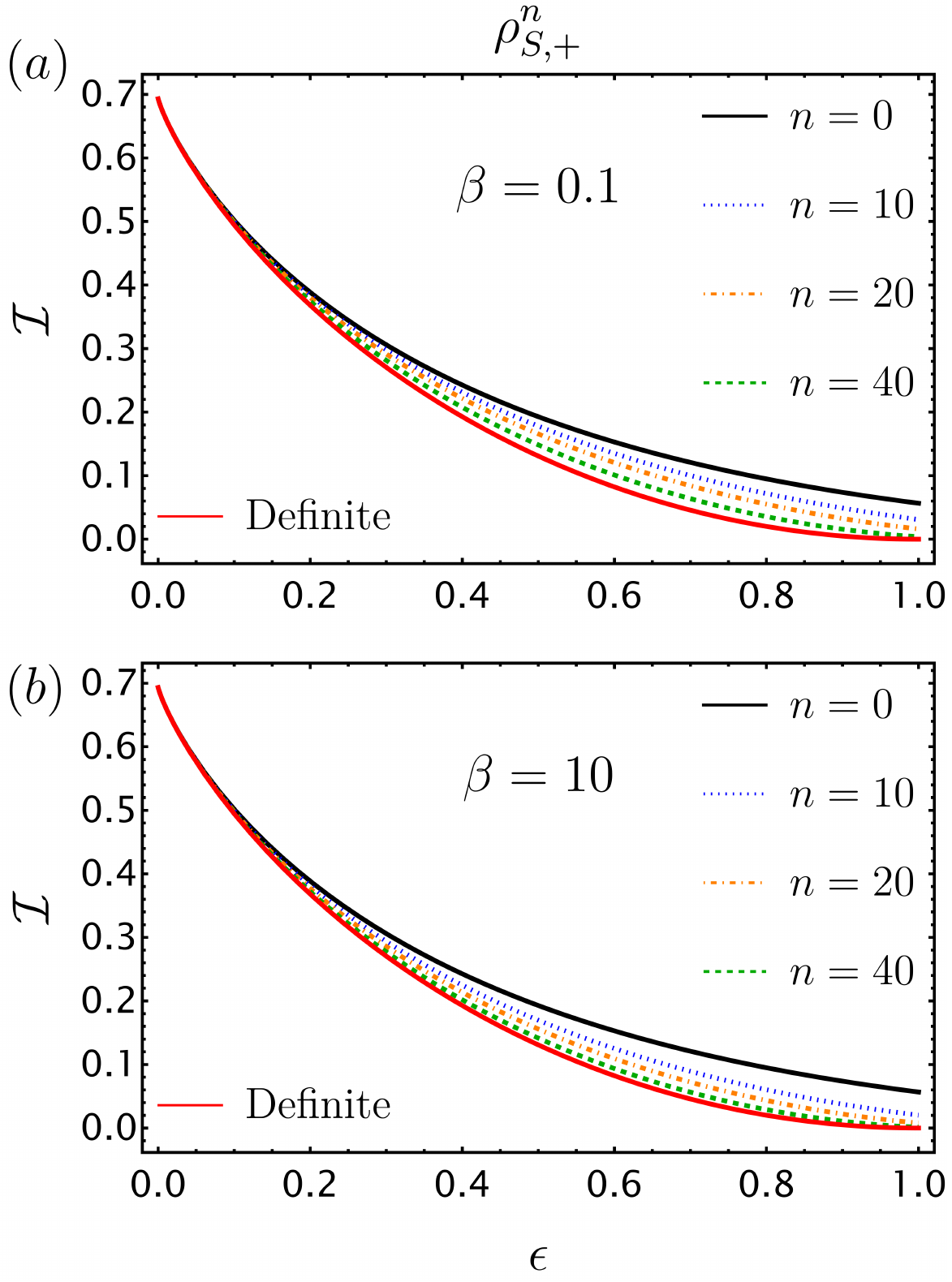}
\caption{\textbf{Available information $\mathcal{I}(\rho_{S,+}^n)$ in the system $S$ after the post-selection of the control.} Here, we assume $\rho_{S}=|+\rangle\langle+|_{S}$, $\omega_{S}=\omega=1$, and $g\tau=0.2$. $\beta$ has units with the inverse of $\omega_S$. We consider the behavior of $\mathcal{I}(\rho_{S,+}^n)$ as a function of the strength $\epsilon$ of the monitoring maps for the case the control has been exposed to an environment with (a) high and (b) low temperatures before its post-selection.}
\label{PlotMonitoramentoPLUS}
\end{figure}

In Fig.~(\ref{PlotMonitoramentoPLUS}) we plot the available information $\mathcal{I}(\rho_{S,+}^n)$---the control is found to be in the $|+\rangle_C$ state---for an increasing number of collisions in two different temperatures: (a) one order of magnitude above and (b) one order of magnitude below the energy scale of the system. In both temperatures, for zero collisions and $\epsilon=1.0$, one sees that the QS followed by post-selection secures some information in the state of the system. As the number of collisions increases, the available information decreases monotonically to the lower limit of definite causal order, where $\epsilon=1.0$ means that no information is left in the state of the system.

\begin{figure}
\center
\includegraphics[width=\columnwidth]{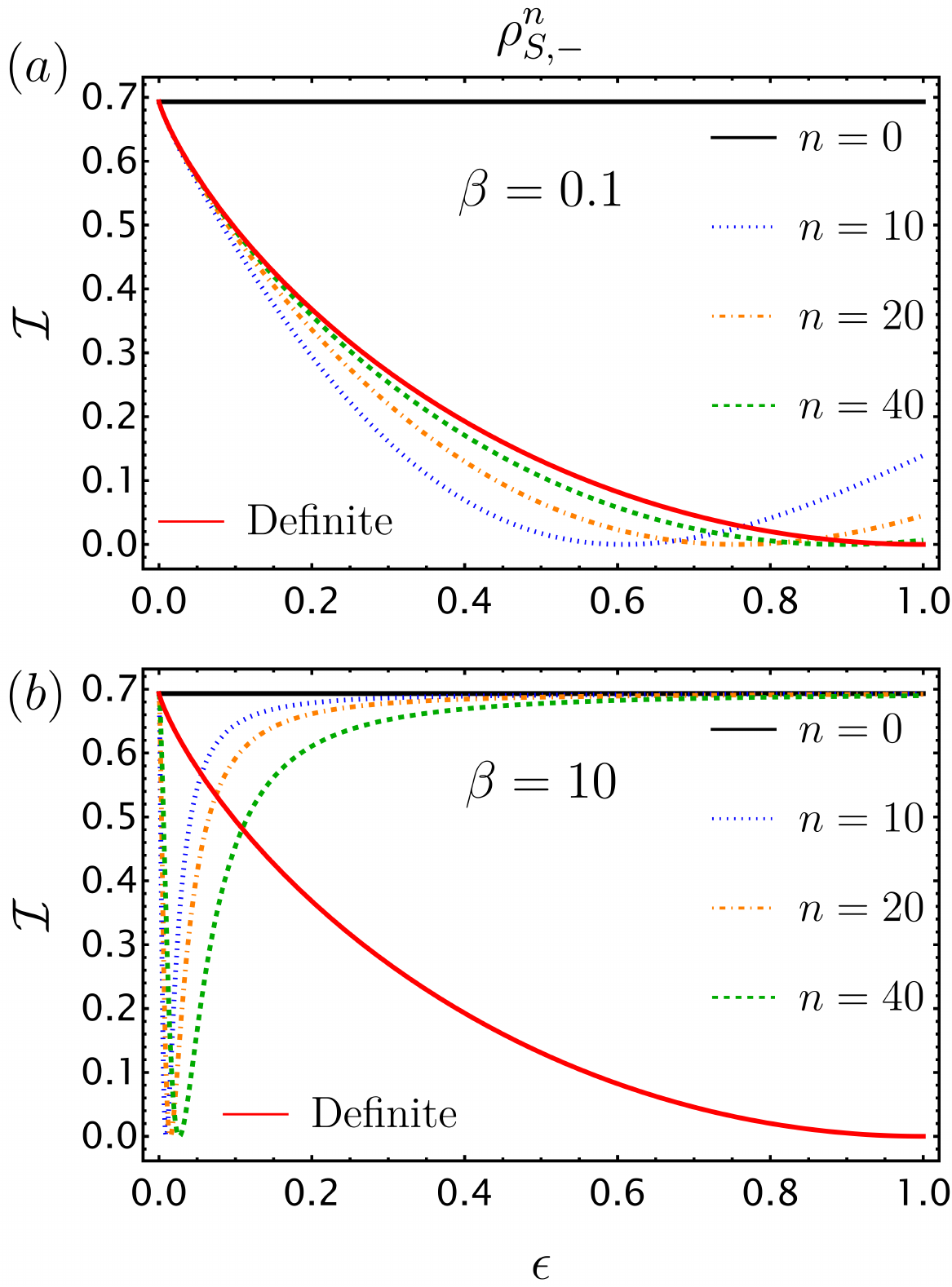}
\caption{\textbf{Available information $\mathcal{I}(\rho_{S,-}^n)$ for (a) high and (b) low temperatures.} Here, we assume $\rho_{S}=|+\rangle\langle+|_{S}$, $\omega_{S}=\omega=1$, and $g\tau=0.2$. $\beta$ has units with the inverse of $\omega_S$. 
As predicted in the Results section, it is clear here that part of the instabilities associated with this post-selection can be attenuated in the low-temperature regime.}
\label{PlotMonitoramentoMINUS}
\end{figure}

On the other hand, when the control is found to be in the $|-\rangle_C$ state the situation changes dramatically. When plotting $\mathcal{I}(\rho_{S,-}^n)$ in Fig.~(\ref{PlotMonitoramentoMINUS}), also for (a) high and (b) low temperatures and different number of collisions, the anticipated consequences previously discussed is observed. Already for high temperatures, when the number of collisions is low ($\rho_{S,-}^0 = |-\rangle \langle -|_S$ exceptionally) the available information is non-monotonic with $\epsilon$, eventually reaching monotonicity for a high number of collisions, for which the available information coincides with the definite order scenario. However, when the temperature is low, the available information has a valley for a small value of $\epsilon$ and grows back to the maximum value for increasing measurement strength. This increase becomes slower for more collisions, such that, for a high number of collisions ($n\gtrsim 300$) we have monotonicity in $\epsilon$ and the curves approximate the definite order behavior. It is worth noting that, following Eq.~\eqref{eq:conditionalstatebdefbindef} we have $\rho_{S,-}^n \equiv |- \rangle \langle -|_S$ for an \textit{arbitrary} $n$. This state has maximum available information $\mathcal{I}(|-\rangle \langle -|_S) = \ln 2$. It means that, as the temperature approaches zero, the probability of projecting the control on $|-\rangle_C$ is suppressed and at the same time the anticipated shielding effect presented in Table~\ref{table:TemperatureQS} is observed. 

\section{QS-based refrigerator}

In the context of quantum thermodynamics, a QS has been employed to design a refrigerator cycle~\cite{Felce2020}. To consider the effect of an external environment, here we present a modified version of the proposed cycle, i.e., it is added a step in which the control interacts with a thermal environment within the collisional model paradigm. Such an interaction takes place right after the switch is performed and before the measurement of the control qubit (see the Supplemental Material for more details).

Let us assume the system $S$ is a qubit described by a Hamiltonian $H_S = -\omega_S \sigma_z^S/2$. Initially, the system is prepared in a thermal state with the inverse temperature $\beta_\text{cold}$ relative to a reference cold bath, i.e., $\Theta_{\beta_\text{cold}} = \text{exp}(-\beta_\text{cold} H_S)/Z_S^\text{cold}$, where $Z_S^\text{cold} = \tr\{ \text{exp}(-\beta_\text{cold} H_S) \}$ is the partition function. The system is then put together to a control auxiliary system $C$ in the ground state of $-\sigma_x$ (i.e., $|+\rangle \langle +|$), which plays the role of the degree-of-freedom conducting the causal order of two identical thermalization maps with cold baths characterized by $\beta_\text{cold}$ (the protocol requires \textit{at least} two baths to operate). Hence, the composite state system-control pre-QS is given by the following product state $\Theta_{\beta_\text{cold}} \otimes |+ \rangle \langle +|_C$.

Then, the QS is applied to the target system according to Eq.~\eqref{eq:switch}, with
\begin{equation}
    M_i = N_j = \sqrt{\frac{\Theta_{\beta_\text{cold}}}{2}} U_i,
\end{equation}
where the $\{U_i\}_i$ form a set of orthogonal unitary operators. Following the procedure in Ref.~\cite{Felce2020}, the state post-QS is given by
\begin{equation}
    \rho_{SC}^0=\frac{1}{2}\left(\Theta_{\beta_{\text{cold}}}\otimes\mathds{1}_{C}+\Theta_{\beta_{\text{cold}}}^{3}\otimes\sigma_{x}\right),
\end{equation}
where the upper index 0 denotes that this is the state pre-open dynamics. 

As a next step, we consider that before measuring $C$ and post-selecting the state of $S$, the control will be interacting with a thermal bath with the inverse of temperature $\beta_{E}$ for a certain time. Here we follow the guidelines presented in the main text, with all the local Hamiltonians and interactions there presented. The final composite system-control state after $n$ collisions is then found to be equal to
\begin{equation}\label{SystemControlStatePostCollision}
    \begin{aligned}
        \rho_{SC}^{n}=&\;\frac{1}{2}\Theta_{\beta_{\text{cold}}}\otimes\left[\mathds{1}_{C}+\left(1-b_{{\rm def}}^{-}(n,f_{E},g\tau)\right)\,\sigma_{x}\right]\\
        &-\frac{1}{2}b_{{\rm indef}}^{-}(n,g\tau)\Theta_{\beta_{\text{cold}}}^{3}\otimes\sigma_{x},
    \end{aligned}
\end{equation}
where $f_{E}\coloneqq\tanh\left(\beta_{\textrm{E}}\omega/2\right)$. Subsequently $C$ is measured in the $\{\ket{+}_{C},\ket{-}_{C}\}$ basis, resulting in two possible procedure branches. On the one hand, if one measures $|+ \rangle \langle +|_C$, the post-selected state of the system $\rho_{S,+}^n$ is classically thermalized to the cold temperature $\beta_\text{cold}$ and the cycle is repeated. On the other hand, in case one measures $|- \rangle \langle -|_C$, the post-selected state of the system $\rho_{S,-}^n$ goes through two consecutive classical thermalizations with a hot and cold bath, respectively characterized by $\beta_{\text{hot}}$ and $\beta_{\text{cold}}$, s.t., $\ensuremath{\beta}_{\text{cold}} > \ensuremath{\beta}_{\text{hot}}$. This last step closes the cycle in this branch which allows the repetition of the whole procedure.

The post-measurement state of the system after $n$ collisions is written as
\begin{equation}\label{SystemPostMeasurement}
    \rho_{S,\pm}^{n}=\frac{\Theta_{\beta_{\text{cold}}}}{2p_{{\rm post}}^{n}(\pm)}\left[b_{{\rm def}}^{\pm}(n,f_{E},g\tau)+b_{{\rm indef}}^{\pm}(n,g\tau)\Theta_{\beta_{\text{cold}}}^{2}\right]
\end{equation}
with measurement probability given by
\begin{equation}
    p_{{\rm post}}^{n}(\pm)=\frac{b_{{\rm def}}^{\pm}(n,f_{E},g\tau)}{2}+\frac{b_{{\rm indef}}^{\pm}(n,g\tau)}{2}\left(1-\frac{3}{4}\text{sech}^{2}\left(\frac{\beta_{cold}\omega_{S}}{2}\right)\right).
\end{equation}
The refrigerator works by effectively removing energy from the cold reservoir in a cyclic manner. The energetic exchange from each step can be straightforwardly computed, as well as its average quantities over many cycles (see Supplemental Material for more details). In this sense, the average heat transferred from the cold bath is given by $\overline{Q}_{n}\equiv p^n_{\rm post}(-)Q_{n,-}$, where
\begin{equation} 
    \begin{split}
        Q_{n,-}\coloneqq&-\omega_{S}\frac{b_{{\rm indef}}^{-}(n,g\tau)}{8p_{{\rm post}}^{n}(-)}\tanh\left(\frac{\ensuremath{\beta}_{\text{cold}}\text{\ensuremath{\omega_{S}}}}{2}\right)\text{sech}^{2}\left(\frac{\ensuremath{\beta}_{\text{cold}}\text{\ensuremath{\omega_{S}}}}{2}\right)\\
        &+\frac{1}{2}\ensuremath{\omega_{S}}\left[\tanh\left(\frac{\ensuremath{\beta}_{\text{hot}}\text{\ensuremath{\omega_{S}}}}{2}\right)-\tanh\left(\frac{\ensuremath{\beta}_{\text{cold}}\text{\ensuremath{\omega_{S}}}}{2}\right)\right]
    \end{split}
\end{equation}
is the heat exchanged from a single cycle of the branch if one measures $|-\rangle\langle-|_C$. 

\begin{figure}
\center
\includegraphics[width=\columnwidth]{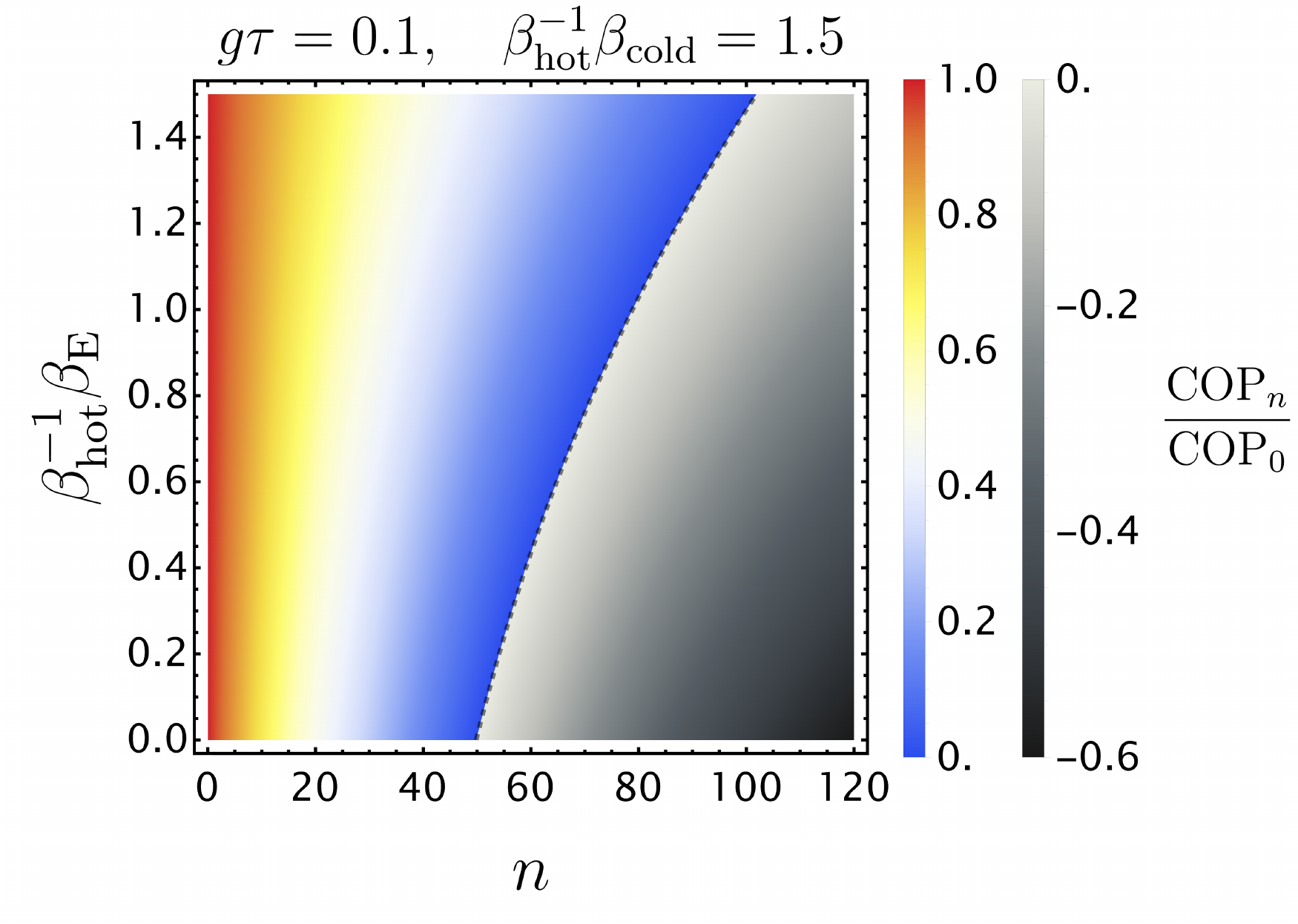}
\caption{\textbf{Normalized coefficient of performance (COP) for $0\leqslant\beta_{E}\leqslant\beta_\text{cold}$ as a function of the number of collisions $n$.} The dashed line represents the point where $\textrm{COP}_{n}/ \textrm{COP}_{0}=0$ and separates the regions where $0\leqslant\textrm{COP}_{n}/ \textrm{COP}_{0}\leqslant1$ (red/blue tones) and $\textrm{COP}_{n}/ \textrm{COP}_{0}<0$ (white/black tones and out of the regime of interest). Parameters: $\omega_{S}=\omega=1$, $\ensuremath{\beta}_{\text{hot}}=1$, $\ensuremath{\beta}_{\text{cold}}=1.5\ensuremath{\beta}_{\text{hot}}$, $g\tau=0.1$ and $\Delta \beta_{E}=0.01$ for the density plot.}
\label{DensityPlot1}
\end{figure}

Additionally, since the average energetic cost of the measurement is null, the average work expended for running the refrigerator is entirely due to Landauer's erasure process \cite{landauer} once one considers the stored measured information, i.e.,
\begin{equation}\label{EnergeticCost}
    \overline{\mathcal{W}}_{n}\coloneqq\mathcal{W}_{n}^{\text{erasure}}\equiv-\frac{1}{\beta_{\text{hot}}}\sum_{k=\pm}p_{{\rm post}}^{n}(k)\ln\left(p_{{\rm post}}^{n}(k)\right).
\end{equation}
Note that we consider the erasure to be performed with the accessible hot bath, so the cold one remains unperturbed during such a process and no other bath is necessary to be included.

Along these lines, the efficiency of the refrigerator can be quantified by the coefficient of performance (COP), defined as the ratio of the heat transferred from the cold bath over the work cost, i.e., 
\begin{equation}
    \textrm{COP}_{n} \coloneqq \frac{\overline{Q}_{n}}{\overline{\mathcal{W}}_{n}}=p^n_{\rm post}(-)\frac{Q_{n,-}}{\mathcal{W}_{n}^\text{erasure}}.
\end{equation}

Fig. \ref{DensityPlot1} shows the refrigerator's performance behavior considering the control is an open system, for different numbers $n$ of collisions and distinct values of $\beta_{E}$ for the external thermal bath, s.t., $0\leqslant\beta_{E}\leqslant\beta_\text{cold}$. As expected, such interaction, characterized by the collisions, decreases the refrigerator's cooling ability, i.e., $\textrm{COP}_{n} < \textrm{COP}_{0}$ for $n>0$. From Eq.~\eqref{SystemControlStatePostCollision} it is clear the composite system-control state asymptotically approaches a causally ordered product state in terms of $n$, such that both states $S$ and $C$ are locally thermal. The open control dynamics destroy the desired correlations, which decreases the amount of heat extracted from the cold bath due to the application of the QS. Also, it is possible to observe the COP decay slower for lower temperatures, closer to the cold bath one, which means the refrigerator functioning is more resilient over low-temperature perturbation.

\begin{figure}
\center
\includegraphics[width=\columnwidth]{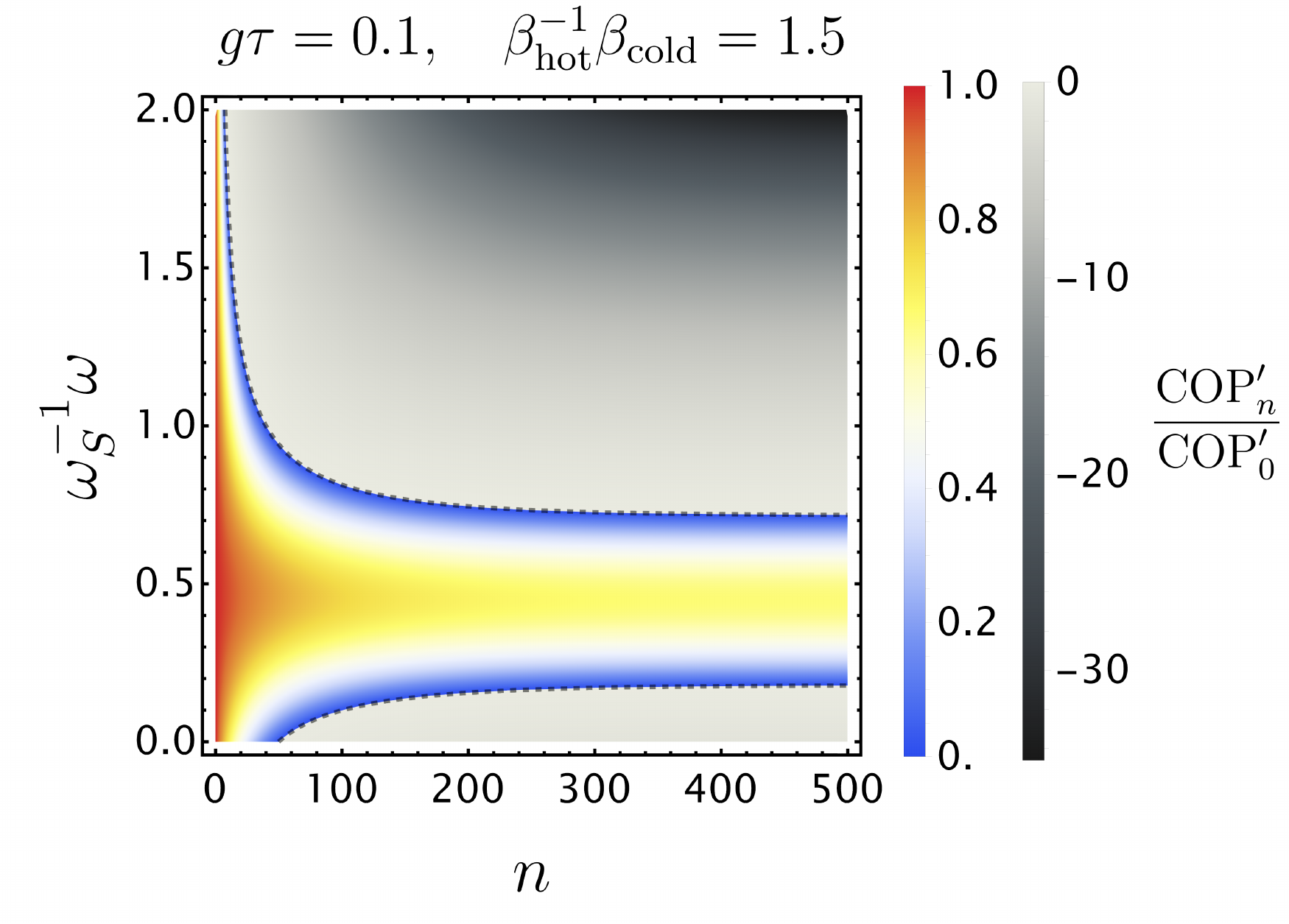}
\caption{\textbf{Normalized coefficient of performance (COP), considering the control heat exchange $q_{n}$, for $0\leqslant\omega\leqslant2\omega_{S}$, as a function of the number of collisions $n$.} The dashed line represents the point where $\textrm{COP}_{n}^{\prime}/ \textrm{COP}_{0}^{\prime}=0$ and separates the regions where $0\leqslant\textrm{COP}^{\prime}_{n}/ \textrm{COP}^{\prime}_{0}\leqslant1$ (red/blue tones) and $\textrm{COP}^{\prime}_{n}/ \textrm{COP}^{\prime}_{0}<0$ (white/black tones and out of the regime of interest). Parameters: $\omega_{S}=1$, $\ensuremath{\beta}_{\text{hot}}=1$, $\ensuremath{\beta}_{\text{cold}}=1.5\ensuremath{\beta}_{\text{hot}}$, $g\tau=0.1$  and $\Delta \omega=0.01$ for the density plot.}
\label{DensityPlot2}
\end{figure}

Nevertheless, if the control is, in fact, into contact with the cold bath, such that $\beta_{E}\equiv\ensuremath{\beta}_{\text{cold}}$, then one should also take into account its energy change. Such a situation is reasonable for settings where the control cannot be fully detached from the other physical systems, particularly from the cold bath under consideration. In this sense, the heat transferred for the control after $n$ collisions with the cold bath is given by
\begin{equation}\label{ControlHeat}
    \begin{split}
        q_{n}=&-\frac{3}{8}\omega\text{sech}^{2}\left(\frac{\ensuremath{\beta}_{\text{cold}}\text{\ensuremath{\omega_{S}}}}{2}\right)+\frac{1}{2}\omega b_{{\rm def}}^{-}(n,f_{\text{cold}},g\tau)\\&+\frac{1}{2}\omega b_{{\rm indef}}^{-}(n,g\tau)\left(1-\frac{3}{4}\text{sech}^{2}\left(\frac{\ensuremath{\beta}_{\text{cold}}\text{\ensuremath{\omega_{S}}}}{2}\right)\right).
    \end{split}
\end{equation}
Hence, both the average heat and COP should be modified, s.t., $\overline{Q^{\prime}}_{n}=p^n_{\rm post}(-)Q_{n,-}+q_{n}$ and
\begin{equation}
    \textrm{COP}_{n}^{\prime}=p^n_{\rm post}(-)\frac{Q_{n,-}}{\mathcal{W}_{n}^\text{erasure}}+\frac{q_{n}}{\mathcal{W}_{n}^\text{erasure}}.
\end{equation}
Note that in such a scenario, one is effectively including $C$ into the working substance of the refrigerator. Fig. \ref{DensityPlot2} shows how the $\textrm{COP}_{n}^{\prime}$ behaves in terms of $\omega$ and $n$ when the control is explicitly considered. In particular, it is possible to see the normalized COP remains positive for a specific gap bandwidth. These values correspond to the parameter region where $C$ can extract energy from the cold bath after the switch application, i.e., under these conditions one guarantees heat flux such that $q_{n}>0$, which assists the cooling process (see Supplemental Material for more details).

\end{appendix}

\acknowledgements{The authors recognize the importance of the Quantum Speedup 2023, organized by the ICTQT, for being the event where the initial talks that led to this work started. O.A.D.M. acknowledges the support from the Foundation for Polish Science (IRAP project, ICTQT, contract no. 2018/MAB/5, co-financed by EU within Smart Growth Operational Programme) and the INAQT (International Network for Acausal Quantum Technologies) Network. A.H.A.M. acknowledges the support from the Polish National Science Centre grant OPUS-21 (No: 2021/41/B/ST2/03207).  R.D.B. acknowledges support by the Digital Horizon Europe project FoQaCiA, Foundations of quantum computational advantage, GA No. 101070558, funded by the European Union. A.C.O.J. acknowledges the support from the QuantERA II Programme (VERIqTAS project) that has received funding from the European Union's Horizon 2020 Research and Innovation Programme under Grant Agreement No 101017733 and from the Polish National Science Center (project No 2021/03/Y/ST2/00175). I.L.P. acknowledges financial support from the ERC Advanced Grant FLQuant. P.R.D. acknowledges support from the NCN Poland, ChistEra-2023/05/Y/ST2/00005 under the project Modern Device Independent Cryptography (MoDIC).}

\bibliography{refs}

\onecolumngrid
\newpage

\eqnumreset

\section*{Supplemental Material}

\subsection*{Refrigerator}

The flowchart describing the main steps of the refrigeration cycle introduced in the Appendices is shown in Fig.~\ref{fig:VedralCycleFlow}.

\subsubsection*{Heat}

\begin{figure*}[b]
    \centering
    \includegraphics[width=0.8\textwidth]{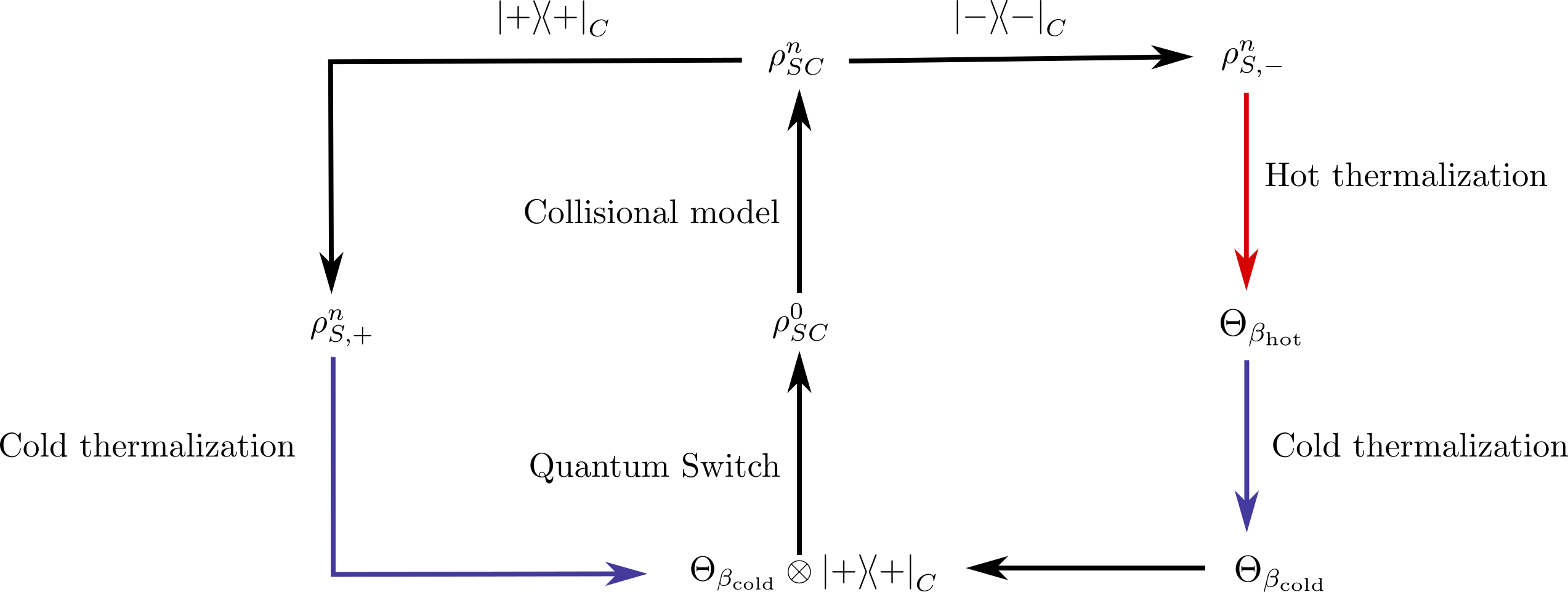}
    \caption{\textbf{Pictorial representation of the refrigeration cycle proposed in Ref.~\cite{Felce2020}.} The system (a qubit) is initialized in the thermal state $\Theta_{\beta_\text{cold}}$ with inverse temperature $\beta_\text{cold}$. An auxiliary system in the state $\ketbra{+}{+}_C$ acting as the control is put in a product state with the system $\Theta_{\beta_\text{cold}}\otimes \ketbra{+}{+}_C$ to start the protocol. After that, the quantum switch is applied, which requires the use of two or more cold baths at $\beta_\text{cold}$. The final state $\rho_{SC}^0$ is then assumed to go through $n$ collisions ($\rho_{SC}^n$) affecting solely the control until the measurement of this subsystem is made in the $\{\ket{+}_{C},\ket{-}_{C}\}$ basis. If the result of the measurement of the control is $\ketbra{+}{+}_C$, then the post-selected state of the system $\rho_{S,+}^n$ is classically thermalized to the cold temperature $\beta_\text{cold}$ and the cycle starts again (positive branch). On the other hand, if after the measurement the control is found to be in the $\ketbra{-}{-}_C$ state, the post-selected state of the system $\rho_{S,-}^n$ goes through two consecutive classical thermalizations: one with a hot bath at $\beta_\text{hot}$ and another one with a cold bath at $\beta_\text{cold}$, closing the cycle (negative branch). We assume that we have a large number of auxiliary systems initially set at the state $\ketbra{+}{+}_C$ to be used as the control at each run of the cycle.}
    \label{fig:VedralCycleFlow}
\end{figure*}

The refrigeration function is achieved whenever the average energy exchanged by the cold bath is positive, i.e., $\overline{Q}_{n}=p^n_{\rm post}(-)Q_{n,-}+p^n_{\rm post}(+)Q_{n,+} > 0$, where $p^n_{\rm post}(\pm)$ is the probability of measuring the state $|\pm\rangle\langle\pm|_C$ and $Q_{n,\pm}$ is the net heat from the positive (negative) branch of Fig.~\ref{fig:VedralCycleFlow}, which consists of the system's energy change due to the measurement of $C$ and the classical thermalization undergone by $S$. Such a condition constrains the parameter-space region for a functioning refrigerator.

On the one hand, for the positive branch one can show the net energy change is null, s.t., 
\begin{equation}
    \begin{split}
        Q_{n,+}&\coloneqq {\rm \tr_{S}}\left[\left(\rho_{S,+}^{n}-\Theta_{\beta_{\text{cold}}}\right)H_{S}\right]+{\rm \tr_{S}}\left[\left(\Theta_{\beta_{\text{cold}}}-\rho_{S,+}^{n}\right)H_{S}\right]\\&=0.
    \end{split}
\end{equation}
On the other hand, for the negative branch one obtains
\begin{equation}
    \begin{split}
       Q_{n,-}&\coloneqq{\rm \tr_{S}}\left[\left(\rho_{S,-}^{n}-\Theta_{\beta_{\text{cold}}}\right)H_{S}\right]+{\rm \tr_{S}}\left[\left(\Theta_{\beta_{\text{cold}}}-\Theta_{\beta_{\text{hot}}}\right)H_{S}\right]\\&=\;-\omega_{S}\frac{b_{{\rm indef}}^{-}(n,g\tau)}{8p_{{\rm post}}^{n}(-)}\tanh\left(\frac{\ensuremath{\beta}_{\text{cold}}\text{\ensuremath{\omega_{S}}}}{2}\right)\text{sech}^{2}\left(\frac{\ensuremath{\beta}_{\text{cold}}\text{\ensuremath{\omega_{S}}}}{2}\right)\\&+\frac{1}{2}\ensuremath{\omega_{S}}\left[\tanh\left(\frac{\ensuremath{\beta}_{\text{hot}}\text{\ensuremath{\omega_{S}}}}{2}\right)-\tanh\left(\frac{\ensuremath{\beta}_{\text{cold}}\text{\ensuremath{\omega_{S}}}}{2}\right)\right]
    \end{split}
\end{equation}
with $\ensuremath{\beta}_{\text{cold}} > \ensuremath{\beta}_{\text{hot}}$. Thus, the average heat transferred from the cold bath becomes
\begin{equation}
    \overline{Q}_{n}=p^n_{\rm post}(-)Q_{n,-}
\end{equation}

\begin{figure}
\center
\includegraphics[width=0.5\textwidth]{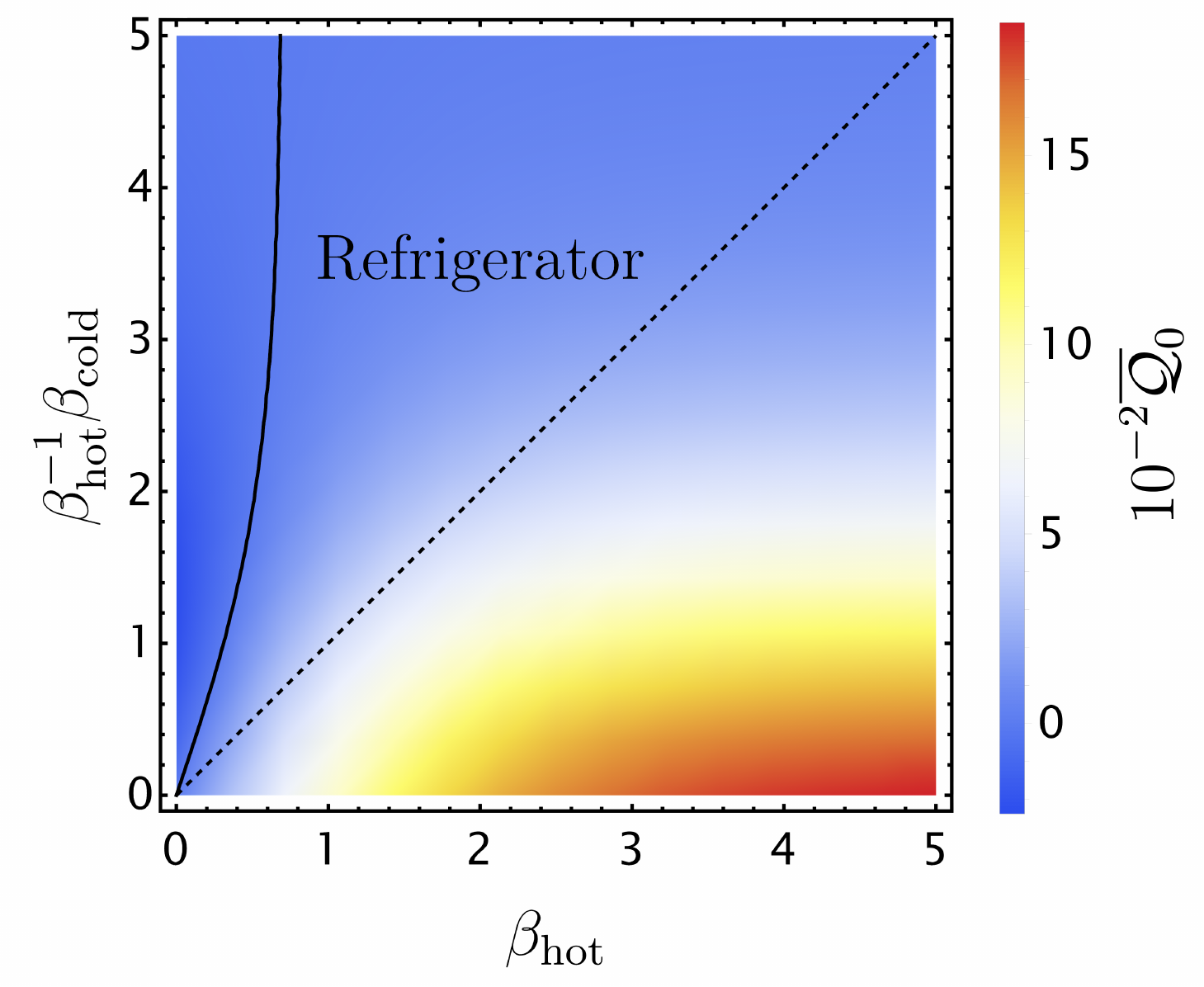}
\caption{\textbf{Refrigeration condition for closed control, i.e., $\overline{Q}_{0}>0$.} The dashed line represents the points where $\ensuremath{\beta}_{\text{cold}} = \ensuremath{\beta}_{\text{hot}}$, while the continuous black line indicates $\overline{Q}_{0}=0$. Parameter: $\omega_{S}=1$.}
\label{RefriCondition}
\end{figure}

Along these lines, for closed control ($n=0$), the refrigeration condition simply becomes
\begin{equation}\label{Condition}
    \overline{Q}_{0}= -\frac{\omega_{S}\left(\tanh\left(\frac{\beta_{\text{cold}}\omega_{S}}{2}\right)-3\tanh\left(\frac{\beta_{\text{hot}}\omega_{S}}{2}\right)\right)}{8(\cosh(\beta_{\text{cold}}\omega_{S})+1)}>0.
\end{equation}
Fig.~\ref{RefriCondition} shows the parameter-space region satisfying the previous condition for $\omega_{S}=1$. The black continuous and dashed lines
highlight the area where $\overline{Q}_{0}>0$ and $\ensuremath{\beta}_{\text{cold}} > \ensuremath{\beta}_{\text{hot}}$ are simultaneously satisfied.

\subsubsection*{Work}

Additionally, the average energetic cost of running such a refrigerator is due to the measurements performed at the control, i.e., by the work expense for measuring $C$, and by Landauer's erasure process of the stored information in a register. The former is given by
\begin{equation}
    \mathcal{W}_{n,\pm}\coloneqq{\rm \tr_{C}\left[\left(|\pm\rangle\langle\pm|_{C}-\rho_{C}^{n}\right)H_{C}\right]}=\frac{\omega}{2}\left(2p_{{\rm post}}^{n}(+)-1\mp1\right),
\end{equation}
such that $p^n_{\rm post}(-)\mathcal{W}_{n,-}+p^n_{\rm post}(+)\mathcal{W}_{n,+}=0$. The latter is defined as 
\begin{equation}
    \mathcal{W}_{n}^\text{erasure} \coloneqq -\frac{1}{\beta_{\text{reg}}}\Delta S_{n,reg}=-\frac{1}{\beta_{\text{reg}}}\sum_{k=\pm}p^n_{\rm post}(k)\ln\left(p^n_{\rm post}(k)\right),
\end{equation}
where $S_{n, reg}$ is the Shannon entropy of the register and $\beta_{\text{reg}}$ is the inverse of the temperature of the bath used to reset the register. Since the thermodynamic cycle consists of two thermal baths we will consider $\beta_{\text{reg}}\equiv\beta_{\text{hot}}$ such that the cold one remains unperturbed during this process. Therefore, the average energetic cost of the refrigerator can be written as Eq.~\eqref{EnergeticCost}.

\subsubsection*{Control heat}

\begin{figure}
    \center
    \includegraphics[width=0.5\textwidth]{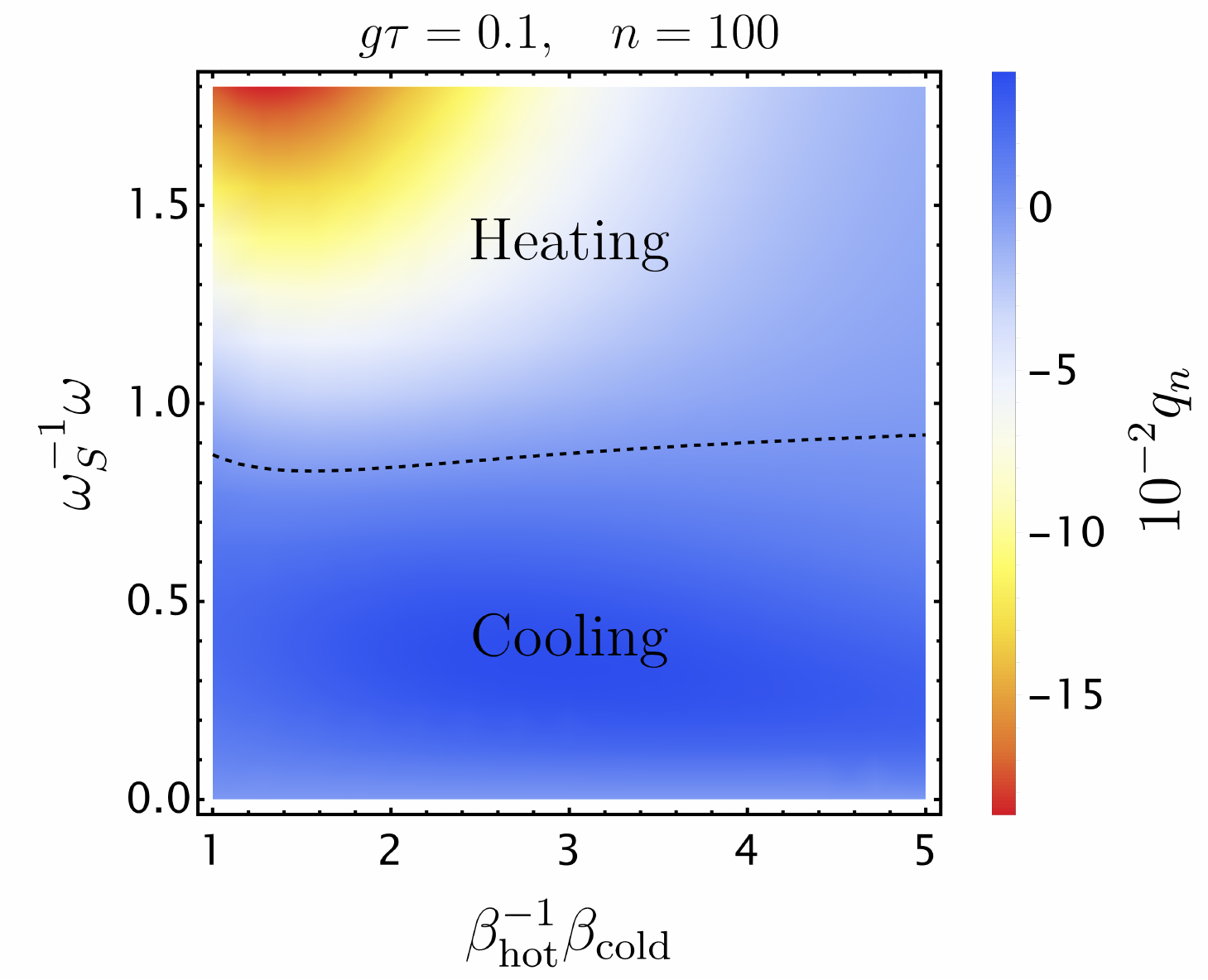}
    \caption{\textbf{Heat exchange between the cold bath and the control system in terms of $\omega$ and $\ensuremath{\beta}_{\text{cold}}$ (Eq.~\eqref{ControlHeat}), for $n=100$ collisions.} The dashed line represents the point where $q_{n}=0$, separating the heating and cooling regions. Parameters: $\omega_{S}=1$, $\ensuremath{\beta}_{\text{hot}}=1$ and $g\tau=0.1$.}
    \label{ControlworkingRegion}
\end{figure}

Eq.~\eqref{ControlHeat} quantifies the amount of heat transferred from the cold bath to the control after $n$ collisions. Fig.~\ref{ControlworkingRegion} shows $q_{n}$ for $n=100$, and highlights the refrigeration working region in terms of $\omega$ and $\ensuremath{\beta}_{\text{hot}}\leqslant\ensuremath{\beta}_{\text{cold}}\leqslant 5$. The cooling area is characterized by $q_{n}>0$, such that heat is extracted from the cold bath.

\end{document}